\pdfoutput=1
\documentclass[letterpaper,abstract=true]{scrartcl}
\usepackage[letterpaper,margin=2.5cm,includefoot]{geometry}
\usepackage[utf8]{inputenc}
\usepackage[english]{babel}
\usepackage[T1]{fontenc}

\usepackage{lmodern}
\usepackage{amsmath,amssymb,mathtools,bm}
\usepackage{csquotes}
\usepackage{amstext}
\usepackage{amsthm}
\usepackage{bbm}
\usepackage{enumerate}
\usepackage{aliascnt}
\usepackage{hyperref}
\usepackage[nameinlink,capitalise]{cleveref}
\usepackage{dsfont}
\usepackage{comment}
\usepackage{color}

\usepackage{fancyvrb}

\makeatletter

\numberwithin{equation}{section}

\newtheorem{thm}{Theorem}[section]

\newaliascnt{lem}{thm}
\newtheorem{lem}[lem]{Lemma}
\aliascntresetthe{lem}

\newaliascnt{prop}{thm}
\newtheorem{prop}[prop]{Proposition}
\aliascntresetthe{prop}

\newaliascnt{cor}{thm}
\newtheorem{cor}[cor]{Corollary}
\aliascntresetthe{cor}

\theoremstyle{definition}

\newaliascnt{defn}{thm}

\aliascntresetthe{defn}

\renewcommand*{\thehyp}{\Alph{hyp}}

\theoremstyle{remark}

\newaliascnt{rem}{thm}
\newtheorem{rem}[rem]{Remark}
\aliascntresetthe{rem}

\newaliascnt{ex}{thm}
\newtheorem{ex}[ex]{Example}
\aliascntresetthe{ex}

\crefname{thm}{Theorem}{Theorems}
\Crefname{thm}{Theorem}{Theorems}

\crefname{lem}{Lemma}{Lemmas}
\Crefname{lem}{Lemma}{Lemmas}

\crefname{prop}{Proposition}{Propositions}
\Crefname{prop}{Proposition}{Propositions}

\crefname{cor}{Corollary}{Corollaries}
\Crefname{cor}{Corollary}{Corollaries}

\crefname{defn}{Definition}{Definitions}
\Crefname{defn}{Definition}{Definitions}

\crefname{rem}{Remark}{Remarks}
\Crefname{rem}{Remark}{Remarks}

\crefname{ex}{Example}{Examples}
\Crefname{ex}{Example}{Examples}

\crefname{hyp}{Hypothesis}{Hypotheses}
\Crefname{hyp}{Hypothesis}{Hypotheses}
\makeatletter
\renewcommand{\@upn}{} \makeatother

\usepackage[inline]{enumitem}
\newlist{enumthm}{enumerate}{1} \setlist[enumthm]{label=\upshape(\roman*),ref=\thethm~(\roman*)}  \crefalias{enumthmi}{thm} \newlist{enumcor}{enumerate}{1}
\setlist[enumcor]{label=\upshape(\roman*),ref=\thecor~(\roman*)}
\crefalias{enumcori}{cor}
\newlist{enumlem}{enumerate}{1}
\setlist[enumlem]{label=\upshape(\roman*),ref=\thelem~(\roman*)}
\crefalias{enumlemi}{lem}
\newlist{enumprop}{enumerate}{1}
\setlist[enumprop]{label=\upshape(\roman*),ref=\theprop~(\roman*)}
\crefalias{enumpropi}{prop}
\newlist{enumhyp}{enumerate}{1}
\setlist[enumhyp]{label=\upshape(\roman*),ref=\thehyp~(\roman*)}
\crefalias{enumhypi}{hyp}
\newlist{enumproof}{enumerate*}{1}
\setlist[enumproof]{label=\upshape(\roman*)}
\newlist{enumdef}{enumerate}{1}
\setlist[enumdef]{label=\upshape(\roman*),ref=\thedefn~(\roman*)}
\crefalias{enumdefi}{defn}
\newlist{enumrem}{enumerate}{1}
\setlist[enumrem]{label=\itshape(\alph*),ref=\therem~(\alph*)}
\crefalias{enumrem}{rem}

\makeatletter
\newcounter{subcreftmpcnt} \newcommand\romansubformat[1]{(\roman{#1})} \def\subcref{\@ifstar\@@subcref\@subcref}
\newcommand\@subcref[2][\romansubformat]{\ifcsname r@#2@cref\endcsname
	\cref@getcounter {#2}{\mylabel}\setcounter{subcreftmpcnt}{\mylabel}\hyperref[#2]{\romansubformat{subcreftmpcnt}}\else ?? \fi}   
\newcommand\@@subcref[2][\romansubformat]{\ifcsname r@#2@cref\endcsname
	\cref@getcounter {#2}{\mylabel}\setcounter{subcreftmpcnt}{\mylabel}\romansubformat{subcreftmpcnt}\else ?? \fi}   
\makeatother

\makeatletter
\DeclareRobustCommand{\crefnosort}[1]{\begingroup\@cref@sortfalse\cref{#1}\endgroup
}
\makeatother

\def\endstepsymbol{$\lozenge$}
\def\endclaimsymbol{$\lozenge$}
\newcounter{proofstep}
\AtBeginEnvironment{proof}{\setcounter{proofstep}{0}}

\crefname{proofstep}{Step}{Steps}
\Crefname{proofstep}{Step}{Steps}
\newcounter{proofclaim}
\AtBeginEnvironment{proof}{\setcounter{proofclaim}{0}}

\crefname{proofclaim}{Claim}{Claims}
\Crefname{proofclaim}{Claim}{Claims}

\newcommand{\cA}{{\mathcal A}}

\newcommand{\cH}{{\mathcal H}}

\newcommand{\cO}{{\mathcal O}}

\newcommand{\cS}{{\mathcal S}}

\newcommand{\BC}{{\mathbb C}}

\newcommand{\BN}{{\mathbb N}}
\newcommand{\BR}{{\mathbb R}}

\newcommand{\BZ}{{\mathbb Z}}

\usepackage{dsfont}

\newcommand{\dsone}{{\mathds 1}}

\usepackage{mathrsfs}

\newcommand{\sfd}{{\mathsf d}}
\newcommand{\sfi}{{\mathsf i}}

\newcommand{\IN}{\BN}\newcommand{\IZ}{\BZ}\newcommand{\IR}{\BR}\newcommand{\IC}{\BC}
\newcommand{\N}{\BN}\newcommand{\R}{\BR}

\renewcommand{\i}{\sfi}\newcommand{\Id}{\dsone} \renewcommand{\d}{\,\sfd}

\renewcommand{\Re}{\operatorname{Re}}\renewcommand{\Im}{\operatorname{Im}}

\newcommand{\ran}{\operatorname{ran}}

\renewcommand{\bar}[1]{\overline{#1}}

\DeclareFontFamily{U}{mathx}{\hyphenchar\font45}
\DeclareFontShape{U}{mathx}{m}{n}{
	<5> <6> <7> <8> <9> <10>
	<10.95> <12> <14.4> <17.28> <20.74> <24.88>
	mathx10
}{}
\DeclareSymbolFont{mathx}{U}{mathx}{m}{n}
\DeclareFontSubstitution{U}{mathx}{m}{n}
\DeclareMathAccent{\widecheck}{0}{mathx}{"71}
\DeclareMathAccent{\wideparen}{0}{mathx}{"75}

\DeclareFontFamily{OMX}{MnSymbolE}{}
\DeclareFontShape{OMX}{MnSymbolE}{m}{n}{
	<-6>  MnSymbolE5
	<6-7>  MnSymbolE6
	<7-8>  MnSymbolE7
	<8-9>  MnSymbolE8
	<9-10> MnSymbolE9
	<10-12> MnSymbolE10
	<12->   MnSymbolE12}{}
\DeclareSymbolFont{mnlargesymbols}{OMX}{MnSymbolE}{m}{n}
\SetSymbolFont{mnlargesymbols}{bold}{OMX}{MnSymbolE}{b}{n}
\DeclareMathDelimiter{\llangle}{\mathopen}{mnlargesymbols}{'164}{mnlargesymbols}{'164}
\DeclareMathDelimiter{\rrangle}{\mathclose}{mnlargesymbols}{'171}{mnlargesymbols}{'171}
\DeclareMathDelimiter{\lsem}{\mathopen}{mnlargesymbols}{'102}{mnlargesymbols}{'102}
\DeclareMathDelimiter{\rsem}{\mathclose}{mnlargesymbols}{'107}{mnlargesymbols}{'107}
\DeclareMathDelimiter{\langlebar}{\mathopen}{mnlargesymbols}{'152}{mnlargesymbols}{'152}
\DeclareMathDelimiter{\ranglebar}{\mathclose}{mnlargesymbols}{'157}{mnlargesymbols}{'157}
\DeclareMathDelimiter{\lWavy}{\mathopen}{mnlargesymbols}{'137}{mnlargesymbols}{'137}
\DeclareMathDelimiter{\rWavy}{\mathopen}{mnlargesymbols}{'137}{mnlargesymbols}{'137}

\newcommand{\abs}[1]{\lvert#1\rvert}\newcommand{\Abs}[1]{\left\lvert#1\right\rvert}

\newcommand{\Norm}[1]{\left\lVert#1\right\rVert}

\newcommand{\nn}[1]{\Norm{#1}}

\renewcommand{\:}{\colon}

\newcommand{\dist}{\operatorname{dist}}

\usepackage{xcolor}\usepackage{cancel}
\definecolor{green}{rgb}{0.0, 0.5, 0.5}
\definecolor{yellow}{rgb}{0.5, 0.5, 0}
\definecolor{lgray}{gray}{0.9}
\definecolor{llgray}{gray}{0.95}
\definecolor{lllgray}{gray}{0.975}

\definecolor{darkcerulean}{rgb}{0.03, 0.27, 0.49} 	\definecolor{darkcoral}{rgb}{0.8, 0.36, 0.27}

\newcommand{\ketbra}[2]{ {|{#1}\rangle}\hspace*{-0.9mm}{\langle{#2}|}  }

\usepackage[backend=biber,style=alphabetic,giveninits=true,maxnames=5,maxalphanames=4, ,url=true,isbn=false,eprint=false,alldates=year]{biblatex}
\DeclareFieldFormat{doilink}{\iffieldundef{doi}{#1}{\href{http://dx.doi.org/\thefield{doi}}{#1}}}

\DeclareBibliographyDriver{article}{\usebibmacro{bibindex}\usebibmacro{begentry}\usebibmacro{author/translator+others}\setunit{\labelnamepunct}\newblock
	\usebibmacro{title}\newunit
	\printlist{language}\newunit\newblock
	\usebibmacro{byauthor}\newunit\newblock
	\usebibmacro{bytranslator+others}\newunit\newblock
	\printfield{version}\newunit\newblock
	\usebibmacro{in:}\printtext[doilink]{\usebibmacro{journal+issuetitle}\newunit
		\usebibmacro{byeditor+others}\newunit
		\usebibmacro{note+pages}}\newunit\newblock
	\iftoggle{bbx:isbn}
	{\printfield{issn}}
	{}\newunit\newblock
	\usebibmacro{doi+eprint+url}\newunit\newblock
	\usebibmacro{addendum+pubstate}\setunit{\bibpagerefpunct}\newblock
	\usebibmacro{pageref}\usebibmacro{finentry}}

\newbibmacro{string+doi}[1]{\iffieldundef{doi}{#1}{\href{https://doi.org/\thefield{doi}}{#1}}}
\DeclareFieldFormat{title}{\usebibmacro{string+doi}{\mkbibemph{#1}}}
\DeclareFieldFormat[inproceedings]{title}{\usebibmacro{string+doi}{\mkbibquote{#1}}}
\DeclareFieldFormat[incollection]{title}{\usebibmacro{string+doi}{\mkbibquote{#1}}}
\DeclareFieldFormat[unpublished]{title}{\usebibmacro{string+doi}{\mkbibquote{#1}}}

\renewbibmacro{in:}{\ifentrytype{article}{}{\printtext{\bibstring{in}\intitlepunct}}}

\newcommand{\cs}[1]{\left< #1 \right>}
\newcommand{\css}[1]{\langle #1 \rangle}

\newcommand{\Dom}{\mathcal{D}}

\usepackage{xcolor}

\DeclareMathOperator{\Tr}{Tr}

\title{On the Lattice \texorpdfstring{$\phi^4$}{phi4} Model:\\ Lieb--Robinson Bounds  and Digital Quantum Simulation}
\author{
	Marius Lemm\thanks{Fachbereich Mathematik, Universität Tübingen, Auf der Morgenstelle 10, 72076 Tübingen, Germany. 
		\\ Email: \href{mailto:marius.lemm@uni-tuebingen.de}{marius.lemm@uni-tuebingen.de}.}
	\and
	Oliver Siebert\thanks{Department of Mathematics, One Shields Ave, University of California, 
		Davis, CA-95616, USA.
		\\ Email: \href{mailto:osiebert@ucdavis.edu}{osiebert@ucdavis.edu}.}
}
\date{\today}
\begin{document}
	\maketitle
	\begin{abstract}
		The lattice $\phi^4$ model is a paradigmatic lattice boson model. Jordan--Lee--Preskill \cite{JLP} and Tong--Albert--McClean--Preskill--Su \cite{Tong} have proposed it as a natural candidate for using digital quantum simulation to understand quantum field theories.
		They also identified that a key mathematical obstacle for analyzing the lattice $\phi^4$ model is that, due to the infinite dimensionality of the local on-site Hilbert space, the standard  Lieb--Robinson bounds  are no longer available. Here, we overcome this obstacle by proving a Lieb--Robinson bound that holds uniformly among all initial states of bounded on-site energy. As an application of our method, we present an improved digital quantum simulation of the lattice $\phi^4$ model whose error bound only depends on the local, not the global energy.
		Our proof is inspired by recent progress on Lieb--Robinson bounds for Bose--Hubbard Hamiltonians, but the structural differences of  the $\phi^4$ model pose new challenges as was already pointed out in \cite{Tong}.
	\end{abstract}
	
	\section{Introduction}
	A major goal of modern quantum many-body physics is to develop rigorous algorithms for the efficient digital quantum simulation of such systems on programmable experimental devices. For systems of qudits (also known as quantum spin systems) such algorithms are well-understood with the state-of-the-art algorithm provided by Haah--Hastings--Kothari--Low \cite{HHKL} in 2018; see also \cite{berry_childs_kothari,BCCKS2017}.
	Progress on the experimental side has been equally remarkable, e.g., programmable Rydberg platforms \cite{weimer2010rydberg,weimer2011digital} have been proposed to simulate lattice gauge theories \cite{celi2020emerging,gonzalez2022hardware}.
	
	Although the $\phi^4$ model is one of the simplest instances of a quantum field theory, it is a particularly challenging instance of the quantum many-body problem. Originally considered more of a toy theory, it turned out to describe the Higgs field. It is standard practice in physics and in mathematics to study its discrete analog, the lattice $\phi^4$ model, which is more amenable to rigorous analysis, often in combination with a continuum limit. 
	In this vein, in 2012, Jordan--Lee--Preskill \cite{JLP} proposed using lattice-based quantum simulation devices, in which interactions can be suitably programmed experimentally, to obtain a \textit{fully quantum} simulation of massive lattice $\phi^4$ theory; see also \cite{hariprakash2025strategies}. Using a finite field grid at each lattice site, they obtained a polynomial-time quantum algorithm for computing scattering probabilities in $\phi^4$ theory and used the total energy and lattice volume to bound the number of field values per site needed to represent the state to a prescribed accuracy.
The dependence on the global energy is unfavorable. A natural remedy would be to use spin-system algorithms such as \cite{HHKL}, but these do not apply directly to bosonic systems, because they crucially rely on \textit{Lieb--Robinson bounds (LRBs)} to spatially decompose the unitary dynamics into parallelizable pieces.
	
	Indeed, the standard proofs of LRBs break down for bosonic lattice models due to their unbounded interactions. An interesting first result was obtained in 2009 by Nachtergaele--Raz--Schlein--Sims \cite{NRSS} who proved LRBs for bosonic lattice Hamiltonians of the form
	\begin{equation}
		\label{eq:nrss hamiltonian}
		H=\sum_{x\sim y} (\phi_x-\phi_y)^2 +\sum_x (\pi_x^2+\phi_x^2+V(\phi_x))
	\end{equation}
	in the case where  $V\in C^1$ with integrable derivative, which allows one to treat it as a perturbation of the quadratic part $\phi_x^2$. These bounds were applied in \cite{nachtergaele2010existence} to prove existence and continuity of the infinite-volume dynamics for such Hamiltonians. The assumption on $V$ does not cover the case $V(\phi_x)=\phi_x^4$ which corresponds to the lattice $\phi^4$ model.

	Next, in 2022, Tong--Albert--McClean--Preskill--Su \cite{Tong} introduced a local truncation scheme for another class of bosonic lattice models, which produced a Lieb--Robinson bound that allowed them to apply the ideas of \cite{HHKL}.
	They cover lattice gauge theories and the Hubbard--Holstein model, in which electrons interact with phonons.
	 A key quantity  that they identified, which they call $r$, is how many powers of the local particle number operator are needed to bound the particle-number non-conserving terms. It is essential that this $r$ is not too large to close suitable differential inequalities. The reason that \cite{Tong} can handle the Hubbard--Holstein Hamiltonian is that its electron--phonon coupling, through which an electron emits or absorbs a single phonon, is linear in the bosonic creation and annihilation operators. It is therefore bounded by the square root ($r=1/2$) of the local phonon number. We quote:
	 \vspace*{4mm}
	 
	\textit{``There are certain models of physical interest
		that do not immediately fit in our framework.
		These include models that involve a quadratic
		coupling between bosonic modes, such as the
		Bose--Hubbard model ($r= 1$ in (6b)) and the discretized $\phi^4$ theory ($r = 2$); our analysis handles
		the case where $r<1$ in (6).'' \cite{Tong}\footnote{Here, the quoted $r=1$ for the Bose--Hubbard Hamiltonian refers to the fact that the hopping term (which is the only part that changes the local boson number and thus affects the truncation) is quadratic in creation and annihilation operators and thus scales as the first power ($r=1$) of the local particle number. For the lattice $\phi^4$ Hamiltonian, the term $\phi_x^4=(a_x+a^\dagger_x)^4 / 4$ is quartic and thus requires $r=2$ powers of the local particle number.}}
			 \vspace*{4mm}
	
	This identifies the Bose--Hubbard and lattice $\phi^4$-models as the next natural frontiers.
	For Bose--Hubbard Hamiltonians ($r=1$), the problem has been solved by now, through the development of Lieb--Robinson bounds for these systems, also in the long-range case \cite{Yin2022,Faupin_LR2022,KVS,kuwahara2024enhanced,Lemm2023info,lemm2025quantum,elgart2026lieb,Carlalemm2026liebrobinsonbounds,rubiliani2026lieb,lemm2026dissipation}.
	Of particular importance was the 2024 work by Kuwahara--Vu--Saito \cite{KVS} who established Lieb--Robinson bounds for initial states of controlled boson density and proposed an efficient digital quantum simulation algorithm for Bose--Hubbard Hamiltonians, resolving the $r=1$ case left open in \cite{Tong}. However, deriving Lieb--Robinson bounds and an efficient quantum simulation algorithm for the lattice $\phi^4$ Hamiltonian ($r=2$) has remained open until now.

	In this paper, we resolve both problems (Lieb--Robinson bounds and efficient digital quantum simulation) for the lattice $\phi^4$ model. Our main result (\Cref{thm:main}) establishes a Lieb--Robinson bound for initial quantum states $\sigma$ satisfying bounded on-site energy moments in the sense that
	\[
	\max_{x\in\Lambda} \cs{h_x^p}_\sigma\leq C_p\qquad \textnormal{ for some even $p>d+1$,}
	\]
	where the on-site energy operator $h_x$ is defined in \eqref{eq:on-site-energy} below and $\cs{\cdot}_\sigma=\mathrm{Tr}(\sigma \cdot)$ denotes the expectation value in the quantum state $\sigma$. We then describe two applications of the propagation bounds:
	\begin{itemize}
		\item In Theorem \ref{sim:thm:main}, we provide an improved digital quantum simulation of the lattice $\phi^4$ model whose error bound only depends on the local, not the global energy.  
		Its gate count is polylogarithmic in the inverse of the precision error. The basic idea for deriving such a result from a Lieb--Robinson bound (as we establish in our main result), which was already present in \cite{Tong,KVS}, is to use the suitable Lieb--Robinson bounds to perform a quantitatively controlled truncation of the local dimension. Then, one maps to qubits and applies the HHKL algorithm \cite{HHKL}. It is imperative to control the dependence of the error on the local dimension, which grows with time.
		\item In Corollary \ref{cor:thermodynamic}, we prove convergence of the finite-volume dynamics to the infinite-volume dynamics for infinite-volume initial states with a uniform bound on the on-site energy. This is a standard (and the original) application of Lieb--Robinson bounds in mathematical physics, though it has fewer precedents for lattice bosons, namely only \cite{nachtergaele2010existence} for the Hamiltonian in \eqref{eq:nrss hamiltonian} and \cite{deuchert2025dynamics} for Bose--Hubbard Hamiltonians. An alternative approach without Lieb--Robinson bounds for the construction of an infinite-volume dynamics  for lattice systems with bounded interactions was given by Buchholz for the resolvent algebra \cite{buchholz2017resolvent}, where he obtained continuity on a smaller invariant subalgebra. 
	\end{itemize}
	This work therefore addresses  two related problems that originate in 2009 \cite{NRSS}, respectively, 2012 \cite{JLP}.	 \vspace*{4mm}

	We close the introduction with an intriguing future direction: In view of recent progress on Lieb--Robinson bounds for many-body systems in continuous space \cite{gebert2020lieb,hinrichs2025lieb,lemm2026lieb_continuum}, it would be very interesting to explore (and then use) propagation bounds for  continuous-space versions of $\phi^4$-theory; see, e.g., \cite{gubinelli2021pde}.
	
	\subsection{Setup}	For the setup we will use the convention of \cite{JLP} with fixed lattice spacing $a=1$.
	For a dimension $d\in\mathbb N$ consider a finite subgraph $\Lambda \subset \mathbb Z^d$ with nearest-neighbor edges
	\[
	E_\Lambda :=
	\bigl\{\{x,y\}\subset
	\Lambda:|x-y|_1=1\bigr\}.
	\]
	We consider the Hilbert space
	\[
	\mathcal H_\Lambda
	=
	L^2(\mathbb R^\Lambda)
	\cong
	\bigotimes_{x\in\Lambda}L^2(\mathbb R).
	\]
	For $X\subseteq\Lambda$ we define $E_X$ and $\cH_X$ analogously.
	
	Let $\phi$ and $\pi$ denote the position and momentum operators on $L^2(\R)$, i.e., $\phi$ is multiplication by the coordinate and $\pi=-\i\partial$, both taken as self-adjoint closures on $C_c^\infty(\R)$. For $x\in\Lambda$, let $\phi_x$ and $\pi_x$ denote the same operators acting on the $x$-th tensor factor of $\cH_\Lambda$ and as the identity elsewhere.
	On the common invariant core $C_c^\infty(\mathbb R^\Lambda)$, they satisfy the canonical commutation relations
	\[
	[\phi_x,\pi_y]
	=
	\i   \delta_{xy}\Id,
	\qquad
	[\phi_x,\phi_y]
	=
	[\pi_x,\pi_y]
	=
	0, \qquad x,y \in \Lambda.
	\]
	
	For a bare mass parameter $m_0\geq0$, and a coupling constant
	$\lambda_0>0$, the lattice $\phi^4$ Hamiltonian with open boundary conditions is now defined in the form sense as
	\begin{equation}\label{eq:phi4-gradient}
		H_\Lambda
		=
		\sum_{x\in\Lambda}
		\left(
		\frac{\pi_x^2}{2}
		+
		\frac{(\nabla^\Lambda\phi)^2_x}{2}
		+
		\frac{m_0^2}{2}\phi_x^2
		+
		\frac{\lambda_0}{4!}\phi_x^4
		\right),
	\end{equation}
	where the discrete gradient is given by
	\[
	(\nabla^\Lambda\phi)^2_x
	:=
	\sum_{j=1}^d
	\bigl((\partial_{j}^{\Lambda}\phi)_x\bigr)^2, \qquad 	(\partial_{j}^{\Lambda}\phi)_x
	:=
	\begin{cases}
		\phi_{x+e_j}-\phi_x
		&: x+e_j\in\Lambda,\\
		0
		&: x+e_j\notin\Lambda.
	\end{cases}
	\]
	More precisely, $H_\Lambda$ is the positive self-adjoint operator
	associated with the corresponding closed quadratic form on
	\[
	\mathcal Q_\Lambda
	=
	\bigcap_{x\in\Lambda}
	\left(
	\Dom(\pi_x)\cap \Dom(\phi_x^2)
	\right).
	\]
	By expanding the squares in the gradient term we can rewrite the Hamiltonian as a sum of positive on-site terms together with the inter-site edge couplings,
	\begin{equation}\label{eq:on-site-decomposition}
		H_\Lambda
		=
		\sum_{x\in\Lambda}h_x
		-
		\sum_{\{x,y\}\in E_\Lambda}\phi_x\phi_y
		-
		|\Lambda| \Id
	\end{equation}
	where
	\begin{equation}\label{eq:on-site-energy}
		h_x
		:=
		\Id
		+
		\left(
		\frac{\pi_x^2}{2}
		+
		\frac{m_0^2+\deg_\Lambda(x)}{2}\phi_x^2
		+
		\frac{\lambda_0}{4!}\phi_x^4
		\right), 
	\end{equation}
	with 	$\deg_\Lambda(x):=	\bigl|\{y\in\Lambda:|x-y|_1=1\}\bigr|$. 
	The added identity in \eqref{eq:on-site-energy} ensures $h_x\geq\Id$, and
	the compensating constant $-|\Lambda|\Id$ preserves the
	original normalization of $H_\Lambda$. These identities are
	understood in the sense of quadratic forms. Note that the on-site energies $h_x$ act on
	a single site each and commute with each other, whereas all coupling between
	different sites is contained in the (unbounded) interactions across the edges $-\phi_x\phi_y$. 
	
	For $t\in\mathbb R$, the time evolution of a bounded observable
	$A\in\mathcal B(\mathcal H_\Lambda)$ will be denoted by
	\[
	\tau_t(A) := 	e^{\i t H_{\Lambda}} A 	e^{-\i t H_{\Lambda}}.
	\]
	
	Let $\mathfrak S_1(\cH_\Lambda)$ be the set of trace-class operators on $\cH_\Lambda$, let $\|\sigma\|_1$ denote the trace-norm of $\sigma \in   \mathfrak S_1(\cH_\Lambda)$, and write $\sigma_t := \tau_{-t}(\sigma)$, $t\in\R$.
	A state is $\sigma\in\mathfrak S_1(\cH_\Lambda)$ with $\sigma\geq0$ and
	$\Tr\sigma=1$. For a positive self-adjoint operator $K$ on $\cH_\Lambda$ and
	$0\leq\sigma\in\mathfrak S_1(\cH_\Lambda)$, we write
	\[
	\cs{K}_\sigma :=
	\Tr\big(\sigma K\big)\in[0,\infty]
	\]
	for the expectation value, which can be infinite. Furthermore, we also write $\nn{\cdot}_2$ for the Hilbert--Schmidt norm. 
	
	For $p\in2\IN$ and a $0 \leq \sigma \in \mathfrak S_1(\cH_\Lambda)$ we will be interested in the local $p$-th moments
	\begin{equation}\label{eq:moment}
		M_p(\sigma) := \max_{x\in\Lambda} \cs{ h_x^p}_\sigma,
	\end{equation}
	which are the central quantity of this paper. They control
	the probability of large field values.
	Since the coupling $-\phi_x\phi_y$ is large only where the fields are large,
	finite moments make its unboundedness harmless with high probability. In contrast to the total energy, $M_p(\sigma)$ is a local quantity and does
	not grow with $|\Lambda|$ for states with uniformly bounded on-site moments. 
	
	For $\emptyset \not= X \subset \Lambda$, we introduce the local algebra
	\begin{align*}
		\cA_X=\mathcal B(\cH_X)\otimes\Id_{\Lambda\setminus X}
	\end{align*}
	and write, for $\emptyset \not= X,Y\subset \IZ^d$ and $R \in \IN_0$,
	\begin{gather*}
		\dist(X,Y) := \min_{x\in X,y\in Y}|x-y|_1, \qquad
		X[R]=\{z\in\IZ^d:\dist(X,\{z\})\leq R\}.
	\end{gather*}

	\subsection{Main result: Lieb--Robinson bound}

	Our first main result and application of the propagation bounds concerns the approximation of time-evolved local observables $\tau_t(A)$ by strictly local observables. 
	\begin{thm}[Local approximation of time-evolved local observables]\label{thm:main}
		For each even $p>d+1$, there are $C_0,C>0$, depending only on
		$d,m_0,\lambda_0,p$, with the following property.
		Let $\emptyset\ne X\subset\Lambda$ and $A\in\cA_X$ with $\nn A \leq 1$, $t\in\R$ and $R\in\IN$ satisfy
		$R\geq C_0(1+|t|)$. 
		\begin{enumerate}[label=(\alph*)]
			\item \label{it:localexpect} Then there is
			$A_R(t)\in\cA_{X[R] \cap \Lambda}$ with $\nn{A_R(t)} \leq 1$, such that for every
			$0\leq\sigma\in\mathfrak S_1(\cH_\Lambda)$ with $\Tr\sigma\leq1$ and
			$M_p(\sigma)<\infty$, 	
			\begin{equation*}
				\big|\Tr\big(\sigma(\tau_t(A)-A_R(t))\big)\big|
				\leq C|X|\sqrt{M_p(\sigma)}\,
				\frac{|t|(1+|t|)^{3p-1}}{R^{p-d-1}}.
			\end{equation*}
			
			\item \label{it:localexpect hs norm}	If, additionally, there is a constant $\kappa\geq1$, depending on $A$ and $p$ such that
			\begin{equation}\label{eq:preserve}
				A\Dom(h_x^{p/2})\subset\Dom(h_x^{p/2}),\qquad
				\|h_x^{p/2}Au\|\leq\kappa\|h_x^{p/2}u\|				
			\end{equation}
			for  all $x\in\Lambda$ and $u\in\Dom(h_x^{p/2})$, then
			\begin{equation*}
				\|(\tau_t(A)-A_R(t))\sigma^{1/2}\|_2
				\leq C_\kappa |X|\sqrt{M_p(\sigma)}\,
				\frac{|t|(1+|t|)^{3p-1}}{R^{p-d-1}},
			\end{equation*}
			where $C_\kappa$ may additionally depend on $\kappa$.
		\end{enumerate}
		Moreover, $A_R(t)$ can be chosen independently of $\Lambda$ if $\Lambda\supseteq X[R+1]$.
	\end{thm}
	The theorem shows that, up to a small error, the time-evolved
	observable $\tau_t(A)$ can be replaced by a strictly local observable supported in the
	$R$-neighborhood of $X$, as long as $R$ grows fast enough compared to $\abs t$. 
	More precisely, the ``light cone shape'' is given by
	\[
	R\gtrsim \abs t^{\frac{3p}{p-d-1}}.
	\]
	For large $p$, it behaves as $R\gtrsim \abs t^3$ and it is an interesting question what the optimal light cone is for the lattice $\phi^4$ model.

	We emphasize that, in contrast to the usual Lieb--Robinson bounds for quantum spin systems, but exactly as in the case of Bose--Hubbard Hamiltonians \cite{Yin2022,Faupin_LR2022,KVS,kuwahara2024enhanced,Lemm2023info,lemm2025quantum,elgart2026lieb,Carlalemm2026liebrobinsonbounds,rubiliani2026lieb,lemm2026dissipation},  the error is \textit{not}
	measured in operator norm but needs to be weighted with the state $\sigma$. This
	reflects the fact that the speed of propagation
	depends on the strength of the fields, which in turn is controlled through the moments
	$M_p(\sigma)$.

	Part \ref{it:localexpect hs norm} is in principle the stronger statement, as 	Part~\ref{it:localexpect} can be derived from \ref{it:localexpect hs norm} through Cauchy--Schwarz. However, Part~\ref{it:localexpect} holds without the additional assumption on the observable \eqref{eq:preserve} and with
	a constant independent of $\kappa$. The fact that one can  remove the assumption on the observable when studying expectation values by exploiting cyclicity of the trace was observed recently in \cite{AHLT} in other models, motivated by a different exploitation of cyclicity in the context of Bose--Hubbard Hamiltonians \cite{kuwahara2024enhanced}.

	\begin{rem}
		We can phrase these propagation bounds in terms of commutators.  	Let $p > d+1$ be even.
		There exists a constant $C_0 = C_0(d,m_0,\lambda_0,p)>0$, and for every $\kappa\geq1$,
		another constant $C = C(d,m_0,\lambda_0,p,\kappa)>0$ such that
		\begin{equation}\label{eq:commutator}
			\cs{\Abs{
					[\tau_t(A),B]}^2
			}_\rho^{1/2}
			\leq
			C|X|\sqrt{M_p(\rho)}\,
			\frac{|t|(1+|t|)^{3p-1}}
			{\dist(X,Y)^{p-d-1}},
		\end{equation}
		for all states $\rho$ with
		$M_p(\rho)<\infty$, for all regions $\emptyset\neq X,Y\subset\Lambda$ and times $t\in\R$ satisfying
		$
		\dist(X,Y)\geq C_0(1+|t|),
		$
		for all $A\in\cA_X$ and $B\in\cA_Y$ with
		$\|A\|,\|B\|\leq1$, where $A$ satisfies \eqref{eq:preserve}
		with constant $\kappa$, and where $B$ satisfies
		$M_p(B\rho B^*)\leq\kappa^2M_p(\rho)$.
		
		This simply follows from \Cref{thm:main} with $R=\lfloor\dist(X,Y)/2\rfloor$,
		which is admissible after enlarging $C_0$. Then $X[R]\cap Y=\emptyset$, so
		$A_R(t)$ commutes with $B$, and
		\begin{align*}
			\|[\tau_t(A),B]\rho^{1/2}\|_2
			&\leq\|(\tau_t(A)-A_R(t))B\rho^{1/2}\|_2+\|(\tau_t(A)-A_R(t))\rho^{1/2}\|_2 \\
			&=\|(\tau_t(A)-A_R(t))(B\rho B^*)^{1/2}\|_2+\|(\tau_t(A)-A_R(t))\rho^{1/2}\|_2,
		\end{align*}
		where we square the norm and reorder the operators in the trace, and subsequently apply the second statement in \Cref{thm:main} to $B\rho B^*$ and $\rho$, and use the assumption
		$M_p(B\rho B^*)\leq\kappa^2M_p(\rho)$. Finally, since $R\geq\dist(X,Y)/3$ and
		$p>d+1$, we may replace $R$ by $\dist(X,Y)$ in the bound at the cost of a constant.
	\end{rem}
	
	We now show that the standard Weyl operators are examples of observables $A,B$ being admissible in \Cref{thm:main} and \eqref{eq:commutator}. 
	The reason is that conjugation with a Weyl operator shifts $\phi_x$ and
	$\pi_x$ by constants, so it changes the on-site energy $h_x$ only by terms
	that are polynomial in $a$ and $b$ and of lower order in $\phi_x$ and
	$\pi_x$. 	The detailed proof is given after \Cref{lem:on-site} below.
	\begin{ex}[Weyl operators]\label{ex:weyl}
		For $x\in\Lambda$ and $a,b\in\R$, consider the unitary local Weyl operators
		\[
		W_x(a,b) :=e^{\i(a\phi_x+b\pi_x)}\in\cA_{\{x\}}
		\]
		which were used in the Lieb--Robinson bounds in  \cite[Section~3]{NRSS}. These operators, as well as their self-adjoint real and imaginary parts $\Re W_x(a,b)$ and $\Im W_x(a,b)$,  are admissible as observables $A$ in \Cref{thm:main}. More precisely,  for every even $p\in\IN$, all of them satisfy \eqref{eq:preserve} with
		\[
		\kappa=C(1+|a|+b^2)^p,
		\]
		where $C\geq1$ depends only on $d,m_0,\lambda_0,p$. 
		
		Moreover, any $B=W_y(a,b)$ is a unitary
		satisfying $M_p(B\rho B^*)\leq\kappa^2M_p(\rho)$
		for every state with finite $M_p(\rho)$. Thus they can be used in \eqref{eq:commutator}.
	\end{ex}

	\subsection{Application 1: Digital quantum simulation}
	\label{sec:simulation results}
	
	Our first application of the propagation bounds is the simulation of time-evolved states. More precisely, for a given input state $\rho$ we want to approximate the time-evolved state by a quantum circuit  consisting of  elementary unitaries, operating on a finite-dimensional quantum system. To this end, we first restrict the Hamiltonian to finitely many occupation levels of the local number operators
	\begin{equation*}
		n_x := \frac{\pi_x^2+\phi_x^2-\Id}{2}, \qquad x \in \Lambda,
	\end{equation*}
	and subsequently approximate the resulting finite-dimensional evolution by a quantum circuit. The moment bounds control the error of the first step, whereas  the explicit harmonic-basis matrix elements and previous simulation results for finite-dimensional systems \cite{HHKL,BCCKS2017} control the cost of the second.
	
	We assume throughout this section and in the proofs in \Cref{sec:simulation} that $m_0^2$ and $\lambda_0$ can be computed to $b \in \IN$ bits in time that scales at most polynomially in $b$. Furthermore, all logarithms in the following simulation bounds are to base $2$.

	The orthogonal projections which restrict the Hamiltonian to the first $K+1 \in \IN$ harmonic levels are given by
	\begin{equation}\label{sim:eq:projectors}
		\Pi_{x,K} := {\mathbf 1}_{[0,K]}(n_x),\qquad
		\Pi_K := \prod_{x\in\Lambda}\Pi_{x,K} .
	\end{equation}
	Furthermore, let $\cH_K := \ran \Pi_K \subset \cH_\Lambda$ denote the finite-dimensional Hilbert space of $K+1$ harmonic levels at each site $x$. We have to encode this space in binary numbers in a qubit register with $q$ qubits per site. If we choose $K +1 = 2^q$ and let $N := |\Lambda|\in\IN$ be the number of sites, then
	\begin{align*} 
		\cH_K \cong (\IC^{K+1})^{\otimes N}  \cong (\IC^2)^{\otimes Nq} .
	\end{align*}
	Extra qubits used for intermediate calculations are called ancillary qubits. We will assume that the initial state $\rho$ is supported only on the first $n_0 +1$ harmonic levels, i.e., it is a density matrix over the Hilbert space $\cH_{n_0}$. 
		\begin{thm}[Quantum simulation]\label{sim:thm:main}
		Set $N=|\Lambda|$ and let $n_0\in\N$,
		$T\ge1$, and $\varepsilon\in(0,1)$. 
		There exist integers $q\ge1$ and $a\ge0$, with
		$K:=2^q-1\ge n_0$ and a unitary circuit $U$, consisting of one- and two-qubit
		gates, acting on
		\[
		\cH_K\otimes(\mathbb C^2)^{\otimes a}
		\cong
		(\mathbb C^2)^{\otimes Nq}
		\otimes(\mathbb C^2)^{\otimes a},
		\]
		such that, for every density matrix $\rho$ on $\cH_\Lambda$
		satisfying $\rho=\Pi_{n_0}\rho\Pi_{n_0}$,
		we have
		\begin{equation}\label{sim:eq:error}
			\left\|
			\Tr_{\mathrm{anc}}\!\left(
			U\bigl(\rho\otimes
			|0\rangle\langle0|^{\otimes a}\bigr)U^*
			\right)
			-e^{-iTH_\Lambda}\rho e^{iTH_\Lambda}
			\right\|_1
			\le\varepsilon,
		\end{equation}
		where we identify $\cH_K$ with its harmonic-basis subspace
		of $\cH_\Lambda$.
 The numbers of
working qubits $Nq$, the number of ancillary qubits $a$ and the total number of elementary one- and two-qubit gates $G$ are bounded by
\[
Nq\leq CN \log(2 n_0 T L),\qquad a\leq CL^c, \qquad G \leq CNT\bigl(n_0^4+T^8+L^4\bigr)L^c,
\]
where $L:=\log\!\bigl(2NT n_0/\varepsilon\bigr)$, 
 $C$ depends only on $d,m_0,\lambda_0$, and $c \in \IN$ depends only on $d$.
	\end{thm}
	The choice of harmonic levels serves the circuit construction as it yields explicit sparse matrices. In this basis, $\phi_x$ changes the occupation by one, so both the on-site quartic term and the inter-site interactions have a fixed number of nonzero entries per row bounded independently of $K$, with explicit formulas. The other two regularizations used in \Cref{sec:reduction} are not sufficient. In fact, a spectral cutoff of $h_x$  as in \eqref{eq:cutoff} is also a valid finite-dimensional approximation but does not provide explicit sparse matrices. The field-amplitude cutoff of the individual field operators as in \eqref{eq:field cutoff Hamiltonian}  would leave the on-site Hilbert spaces infinite-dimensional.
	
	The condition on the initial state $\rho$ resembles the assumption of Tong et al.~\cite{Tong}. For the Hubbard--Holstein model, they obtain $\cO(NT(\sqrt{n_0}+T))$ gates~\cite[Appendix E.2]{Tong} up to logarithmic factors, but their truncation theorem does not cover interacting $\phi^4$ theory. Earlier $\phi^4$ algorithms include the scattering algorithms of Jordan--Lee--Preskill \cite{JLP}. Our circuit construction follows the finite-dimensional argument of Kuwahara--Vu--Saito \cite[Methods, Eqs.~(34)--(45)]{KVS}, which combines the spatial decomposition of Haah--Hastings--Kothari--Low \cite{HHKL} with sparse-matrix simulation while keeping the cutoff dependence explicit.
	
	\subsection{Application 2: Existence of the thermodynamic limit of dynamics}
	
	As a second application, we construct the dynamics of the infinite system for general locally normal states with finite local energy moments. To this end, let 
	\[
	\cA:=\overline{\bigcup_{V\subset\mathbb Z^d \text{ finite}}\cA_V}^{\,\|\cdot\|}
	\]
	be the quasi-local $C^*$-algebra. For finite $\Lambda\subset\mathbb Z^d$, we now track the $\Lambda$-dependence of the dynamics and write $\tau_t^\Lambda(B):=e^{\i tH_\Lambda}Be^{-\i tH_\Lambda}$, which extends to a $*$-automorphism of $\cA$. A state $\omega$ on $\cA$ is called \emph{locally normal} if, for every finite $V\subset\mathbb Z^d$, there is a density matrix $\rho_V$ on $\cH_V$ with $\omega(B)=\Tr(\rho_VB)$ for $B\in\cA_V$. We can extend the notion of local energy moments \eqref{eq:moment} to $\omega$ by setting
	\[
	M_p(\omega):=\sup_{\substack{V\subset\mathbb Z^d\text{ finite}\\ x\in V}}\omega\big((h_x^V)^p\big),
	\]
	with $h_x^V$ being the on-site energy \eqref{eq:on-site-energy} restricted to volume $V$ (which only affects the vertex degree).
	
	\begin{cor}[Thermodynamic limit]\label{cor:thermodynamic}
		Let $p>d+1$ be even and let $\omega$ be locally normal with $M_p(\omega) <\infty$. For every $t\in\R$ and $A\in\cA$, the limit
		\[
		\omega_t(A):=\lim_{\Lambda\uparrow\mathbb Z^d}\omega\big(\tau_t^\Lambda(A)\big)
		\]
		along the net of finite subsets of $\mathbb Z^d$, ordered by inclusion, exists, and $\omega_t$ is a state on $\cA$. Furthermore, $t\mapsto\omega_t(A)$ is continuous for every $A\in\cA$.
	\end{cor}
	The proof is given in \Cref{app:thermodynamic}.
	\begin{rem}\label{rem:thermodynamic}
		One can also show that the state $\omega_t$ is still locally normal and  the dynamics acts as a group, i.e.,  $\omega_0=\omega$ and $(\omega_t)_s=\omega_{t+s}$ for all $s,t\in\R$.
		Furthermore, both the moment propagation bounds and the local approximation of time-evolved observables carry over to $\omega$ in the sense that
		\begin{align*}
			M_p(\omega_t)\leq C(1+|t|)^{4p}M_p(\omega)
		\end{align*}
		and  for $\emptyset\neq X\Subset\mathbb Z^d$, $A\in\cA_X$ with $\nn A\leq1$, $t\in\R$, integers $R\geq C_0(1+|t|)$ and finite $\Lambda\supseteq X[R+1]$,
		\[
		\big|\omega_t(A)-\omega\big(\tau_t^\Lambda(A)\big)\big|
		\leq C|X|\sqrt{M_p(\omega)}\,\frac{|t|(1+|t|)^{3p-1}}{R^{p-d-1}}.
		\]
	\end{rem}

	\subsection{Organization of the paper} In Section \ref{sec:reduction}, we prove the main result, Theorem \ref{thm:main}. In Section \ref{sec:simulation}, we establish the application to quantum simulation, Theorem \ref{sim:thm:main}. Finally, in the short Section \ref{app:thermodynamic}, we prove Corollary \ref{cor:thermodynamic}.
	
	\section{Proof of main result}\label{sec:reduction}

	In this part, we prove \Cref{thm:main}, whose proof consists of three steps. First, we show that the local moments $M_p(\sigma_t)$ grow at
	most polynomially in time, uniformly in $\Lambda$ (\Cref{prop:moment}). This
	combines on-site estimates for the anharmonic oscillator $h_x$
	(\Cref{lem:on-site}) with a differential inequality for $t\mapsto M_p(\sigma_t)$,
	which we first derive for a regularized dynamics
	(\Cref{lem:spectral approx}). Second, we use the moment
	bound to replace the couplings on all edges touching $\Omega:=X[R] \cap \Lambda$ by
	bounded couplings with a field-amplitude cutoff (\Cref{lem:clip}). Third, we remove the cutoff
	couplings between $\Omega$ and its complement with standard
	Lieb--Robinson bounds for bounded interactions (\Cref{lem:boundary}), and
	optimize over the cut-off threshold $Q$. Throughout the proof, constants $C$ may change from
	line to line. They depend only on $d,m_0,\lambda_0,p$, but not on the system size $\Lambda$, except in \Cref{lem:spectral approx} where we write $C_\Lambda$.

	\subsection{Moment propagation bounds} \label{sec:moments}
	
	In this section we prove the following propagation bounds on the moments. Note that conservation of energy alone controls only sums over all sites and
	would lead to bounds growing with $|\Lambda|$. Instead, we have to control the moments at every single site
	uniformly in $\Lambda$.
	\begin{prop}[Propagation of moments]\label{prop:moment}
		There is $C\geq1$, depending only on $d,m_0,\lambda_0$, such that for every even $p\in\IN$, every $0\leq\sigma\in\mathfrak S_1(\cH_\Lambda)$ with $M_p(\sigma)<\infty$ and every $t\in\R$,
		\[
		M_p(\sigma_t)^{1/(4p)}\leq C\Bigl(M_p(\sigma)^{1/(4p)}+\left(\sqrt {p/2}+|t|\right)(\Tr\sigma)^{1/(4p)}\Bigr).
		\]
	\end{prop}
	Since the on-site
	energies commute with each other, $h_x^p$ changes in time only through the
	couplings $\phi_x\phi_y$ to the neighbors $y$ of $x$, and this change is of
	lower order. 
	\begin{lem}[On-site bounds]\label{lem:on-site}
		Fix $x\in\Lambda$ and regard the on-site operator $h_x$ as an
		operator on $L^2(\mathbb R)$, i.e., 
		\[
		h:=h_x
		=
		\Id
		+
		\frac{\pi^2}{2}
		+
		\frac{m_0^2+\deg_\Lambda(x)}{2}\phi^2
		+
		\frac{\lambda_0}{4!}\phi^4.
		\]
		There are $C,D\geq1$, depending only on $d,m_0,\lambda_0$, such that
		for every $k\in\IN$ the following hold.
		\begin{enumerate}[label=(\alph*)]
			\item \label{it:compact resolvent}
			The operator $h$ has compact resolvent, and all its eigenfunctions
			belong to $\mathcal S(\mathbb R)$.
			\item \label{it:word estimate}
			Set $\widetilde h=h+Dk^2\Id$.
				Let $\mathcal W_m$ be the set of products of operators $\phi,\pi$, including the empty product $\Id$, of weight at most $m$,
			with weights $1,2$ for $\phi,\pi$, respectively. For $0\leq j\leq k$ and $q=1,2$, we have
			\begin{equation*}
				\|\widetilde h^j w u\|\leq C\|\widetilde h^{j+q/2}u\|,\qquad w\in\mathcal W_{2q},\quad u\in\cS(\R).
			\end{equation*}

			\item \label{it:shifted}
		We have 
			\begin{equation*}
			    \|[\widetilde h^k,\phi]u\|\leq Ck\|\widetilde h^{k-1/2}u\|,\qquad u\in\cS(\R).
			\end{equation*}

			\item \label{it:moment bounds}
			For $w = \phi^4$ and for $w=\tfrac12(\pi^2+\phi^2+\Id)$, we have
			$\Dom(h^k)\subset\Dom(w^{k})$ and 
			\begin{equation*}
			    \|w^{k}u\| \leq C^k\bigl(\|h^ku\|+k^{2k}\|u\|\bigr),
			    \qquad u\in\Dom(h^k).
			\end{equation*}
		\end{enumerate}
	\end{lem}
	\Cref{lem:on-site} collects what we need about a single anharmonic
	oscillator. For large energies, $h\approx\pi^2/2+\lambda_0\phi^4/24$, so
	$\phi$ is of size $h^{1/4}$ and $\pi$ is of size $h^{1/2}$. \Cref{lem:on-site} \ref{it:shifted} says that commuting with $\phi$ cancels one $\pi$ and therefore lowers the size by a factor $h^{1/2}$.
	The shift in $\widetilde h$ is required in order to get constants uniform in the moment order $k$. 

	\begin{proof}
		Write
		\[
		V_x(s):=1+\frac{m_0^2+\deg_\Lambda(x)}{2}s^2
		+\frac{\lambda_0}{24}s^4,
		\qquad h=-\frac12\partial_s^2+V_x.
		\]
		
		\begin{enumerate}[label=(\alph*),wide,nosep]
			\item
			Note that the potential
			$V_x$ is continuous, bounded below, and tends to
			$+\infty$ as $|s|\to\infty$. Hence $h$ has compact resolvent
			by \cite[Theorem~XIII.67]{RS4}.
			
			Let $h\psi=E\psi$. Since $V_x$ is smooth, iteration of the eigenvalue equation
			$\psi''=2(V_x-E)\psi$ shows that $\psi\in H^n_{\mathrm{loc}}(\R)$
			for every $n$, and hence $\psi\in C^\infty(\R)$. Moreover, by
			\cite[Theorem~C.3.3]{simon1982semigroups},
			$
			|\psi(s)|\leq C_a e^{-a|s|}
			$
			for every $a>0$. 
			As $V_x$ is a polynomial, the eigenvalue equation implies that $\psi''$ also decays
			faster than every power. Writing $\psi'$ as an integral over $\psi''$ gives the same decay for $\psi'$.
			By repeated differentiation of the eigenvalue equation
			we get this property for every derivative.
			Hence $\psi\in\mathcal S(\R)$.
			
			\item
		We prove this by induction over $j$ for a constant $D$ chosen below that depends only on $d,m_0,\lambda_0$.
			For $j=0$ and $q=1$, this follows from the form bounds $\pi^2 \leq 2h$ and $\phi^4 \leq 	(4! / \lambda_0) h$.
For $q=2$, move the $\pi$ factors to the right using $[\phi,\pi]=\i\Id$.
A word of weight at most $4$ can have at most four $\phi$'s if it contains
no $\pi$, and at most two $\phi$'s if it contains one $\pi$.
All these words can be controlled by $V_x$ as
\[
\|\phi^r u\|\leq C\|V_xu\| \quad \text{ for } 0\leq r\leq4, \qquad
\|\phi^r\pi u\|\leq C\|V_x^{1/2}\pi u\| \quad  \text{ for }  0\leq r\leq2.
\]
The only word with two momenta is $\pi^2$.
To bound all three norms, we expand $\|hu\|^2$ and write
\[
\|V_xu\|^2+\|V_x^{1/2}\pi u\|^2+\tfrac14\|\pi^2u\|^2
=\|hu\|^2+\tfrac12\langle u,V_x''u\rangle.
\]
Finally, we use $V_x''\leq C V_x\leq C h$ uniformly in $x,\Lambda$  to see that the right-hand side is bounded by $C\|\widetilde h u\|^2$. 

Let $C_{\text{b},0}$ denote the constant $C$ of this base case, which depends only on $d,m_0,\lambda_0$. We will prove the current statement \ref{it:word estimate} with the constant $C_{\text{b}}:=2C_{\text{b},0}$.

For the induction step assume the bounds hold up to $j-1$, with $1\leq j\leq k$. Write
\[
\widetilde h^j w
=w\widetilde h^j+
\sum_{\ell=0}^{j-1}
\widetilde h^\ell[\widetilde h,w]\widetilde h^{j-1-\ell}.
\]
The base case controls the first term $w\widetilde h^j u$ with the constant $C_{\text{b},0} = C_{\text{b}}/2$.
In the sum, commuting with $\widetilde h$ replaces one $\phi$ by
$-\i\pi$ or one $\pi$ by $\i V_x'(\phi)$, which is of order $3$. Thus, each new word gains at most one unit of weight.

First let $w\in\mathcal W_2$. The new words have weight at most $3$,
so the $\mathcal W_4$ bound applies at every $\ell<j$. This yields, for a constant $C$ only depending on $d,m_0,\lambda_0$,
\[
\|[\widetilde h^j,w]u\|
\leq  C jC_{\text{b}}\|\widetilde h^j u\|
\leq C \frac{C_{\text{b}}}{\sqrt D}\|\widetilde h^{j+1/2}u\|,
\]
where in last step we used $\widetilde h^{1/2}\geq\sqrt D\,k\Id \geq\sqrt D\,j\Id$. 
Choosing $\sqrt D\geq 2C$  the sum of commutators is bounded by $C_{\text{b}}/2 \|\widetilde h^{j+1/2}u\|$, proving the $\mathcal W_2$ bound for $j$.

Now let $w\in\mathcal W_4$. Each new word has weight at most $5$.
Removing its last factor $X\in\{\phi,\pi\}$ leaves a word
$U\in\mathcal W_4$. We can then use the $\mathcal W_4$ bound for $\ell<j$, and subsequently
the $\mathcal W_2$ bound just proved for $\ell+1\leq j$ to obtain
\[
\|\widetilde h^\ell UXv\|
\leq C_{\text{b}}\|\widetilde h^{\ell+1}Xv\|
\leq C_{\text{b}}^2\|\widetilde h^{\ell+3/2}v\|.
\]
The same commutator expansion as above then yields
\[
\|[\widetilde h^j,w]u\|
\leq C j C_{\text{b}}^2\|\widetilde h^{j+1/2}u\|
\leq \tfrac{C_{\text{b}}}2\|\widetilde h^{j+1}u\|,
\]
by enlarging $D$ such that $\sqrt D \geq 2CC_{\text{b}}$. This finishes the induction and proves the assertion.

\item We use the same commutator expansion as before and apply  \ref{it:word estimate} with $q=1$ and $w=\pi$ to see that
\[
\|[\widetilde h^k,\phi]u\|
\leq \sum_{\ell=0}^{k-1}
\|\widetilde h^\ell\pi\widetilde h^{k-1-\ell}u\|
\leq kC_{\text{b}}\|\widetilde h^{k-1/2}u\|.
\]

			\item
			Both choices for $w$ are linear
			combinations of products of weight at most $4$, hence,  $\|\widetilde h^j w u\|\leq 2C_{\text{b}}\|\widetilde h^{j+1}u\|$  for $u\in \cS(\R)$ by  \ref{it:word estimate}. Applying this successively for
			$j=0,\ldots,k-1$ yields $\|w^ku\|\leq(2C_{\text{b}})^k\|\widetilde h^ku\|$. 
	To estimate the last norm, expand $u=\sum_n c_n e_n$
	in an orthonormal eigenbasis of $h$, with $he_n=E_ne_n$.
	Since $(E_n+a)^k\leq 2^{k-1}(E_n^k+a^k)$ for each $n$ and $a \geq0$, we have
	\begin{align*}
		\|\widetilde h^ku\|^2
		&= \sum_n (E_n+Dk^2)^{2k}|c_n|^2 
		  \leq  2^{2(k-1)} \sum_n |(E_n^k+(Dk^2)^k)c_n|^2
		= 2^{2(k-1)}\|h^ku+(Dk^2)^ku\|^2 
		\\ &\leq2^{2(k-1)} \bigl(\|h^ku\|+D^k k^{2k}\|u\|\bigr)^2.
	\end{align*}
	By \ref{it:compact resolvent}, the eigenfunction span lies in $\cS(\R)$
	and is a core for $h^k$. Since $w^k$ is closed, the estimate
	extends to all $u\in\Dom(h^k)$.
	 \qedhere
		\end{enumerate}
	\end{proof}
	
	\begin{proof}[Proof of \Cref{ex:weyl}]
	 Set $k=p/2$, and $M=1+|a|+b^2$.
	  The Weyl operator preserves $\cS(\R)$ and shifts $\pi$ by $a$ and $\phi$
	  by $-b$ under conjugation, so
	  \begin{equation}\label{eq:Weyl shift}
	    h_{a,b}:=W_x(a,b)^*hW_x(a,b)=\tfrac12(\pi+a)^2+V_x(\phi-b).
	  \end{equation}
	  Each term in $h_{a,b}$ is a product of weight at most $4$ with coefficient
	  bounded by $CM^2$. Hence, \Cref{lem:on-site} \ref{it:word estimate} with $q=2$ implies
	  $\|\widetilde h^jh_{a,b}v\|\leq CM^2\|\widetilde h^{j+1}v\|$ for $\widetilde h=h+Dk^2\Id$.
	  Iterating for $j=0,\ldots,k-1$ yields
	  \[
	    \|h^kW_x(a,b)u\|=\|h_{a,b}^ku\|\leq(CM^2)^k\|\widetilde h^ku\|
	    \leq C_p M^p\|h^ku\|,
	  \]
	  for all $ u\in\cS(\R)$,
	  since $h\geq\Id$ and $2k=p$. As $\cS(\R)$ is a core for $h^k$, this extends to $\Dom(h^k)$.
	  The same bound holds for $W_x(a,b)^*=W_x(-a,-b)$, hence for $\Re W_x(a,b)$ and $\Im W_x(a,b)$.
	  At every other site $W_x(a,b)$ commutes with the on-site energy. This proves
	  \eqref{eq:preserve}.
	  
	  To verify the hypothesis on $B = W_x(a,b)$ in \eqref{eq:commutator},  write $\rho=\sum_n r_n|\psi_n\rangle\langle\psi_n|$ with $r_n>0$.
	  Since $M_p(\rho)<\infty$, each $\psi_n \in\Dom(h_z^{p/2})$
	  for every site $z$. At $z=x$ we apply \eqref{eq:preserve} and at $z\neq x$,
	  $B$ commutes with $h_z$ and is unitary. Thus, for every $z$,
	  \[
	  \Tr(B\rho B^*h_z^p)
	  =\sum_n r_n\|h_z^{p/2}B\psi_n\|^2
	  \leq \kappa^2\sum_n r_n\|h_z^{p/2}\psi_n\|^2
	  =\kappa^2\Tr(\rho h_z^p).
	  \]
	  Taking the maximum over $z$ shows $M_p(B\rho B^*)\leq\kappa^2M_p(\rho)$.
	\end{proof}
	To prove \Cref{prop:moment}, we would like to differentiate
	$t\mapsto\Tr(\sigma_t h_x^p)$ and estimate the derivative. A priori, it is not clear that this function is finite or differentiable, since $h_x^p$ is unbounded. We therefore first work with a regularized dynamics on a
	finite-dimensional subspace, where all these quantities are finite and
	smooth, and remove the regularization at the end. On $\cH_\Lambda$, we introduce an energy cutoff in the two-site interaction by
	\begin{equation}\label{eq:cutoff} P_L=\prod_{x\in\Lambda} {\mathbf 1}_{[1,L]}(h_x),\qquad
		H^{(L)}= \sum_{x\in\Lambda} h_x - P_L\sum_{\{x,y\}\in E_\Lambda}\phi_x\phi_yP_L 	-
		|\Lambda| \Id,\qquad L\in\IN.
	\end{equation}
	Since $P_L$ commutes with
	every $h_x$, the commutator $[h_x^k,H^{(L)}]$ has the same structure as
	$[h_x^k,H_\Lambda]$, only sandwiched between two factors $P_L$. Hence the
	regularization does not interfere with the moment estimates, and all bounds
	below are uniform in $L$.
	Note that $h_x$ has only discrete eigenvalues by \Cref{lem:on-site} \ref{it:compact resolvent}, so $P_L$ has finite rank for all $L \geq 1$. 
	We write for a $0\leq\sigma\in\mathfrak S_1(\cH_\Lambda)$
	\[
	\sigma_{L,t} :=e^{-itH^{(L)}}P_L\sigma P_Le^{itH^{(L)}}
	\in\mathfrak S_1(\cH_\Lambda),\qquad t\in\R.
	\]
	Furthermore, let \[
	H_{\mathrm{on}} := \sum_{x \in \Lambda} h_x -
	|\Lambda| \Id
	\]
	denote the self-adjoint on-site contribution of the Hamiltonian.
	
	We first show a lower semicontinuity property. It is all we
	need, since we only want upper bounds on the moments of $\sigma_t$, so any bound
	on $\cs{h_x^p}_{\sigma_{L,t}}$ that is uniform in $L$ carries over to
	$\cs{h_x^p}_{\sigma_t}$.
	\begin{lem}[Spectral approximation]\label{lem:spectral approx}
		For each $L \geq 1$, $H^{(L)}$ is self-adjoint on the common domain $\Dom(H_\Lambda) = \Dom(H_{\mathrm{on}})$, its dynamics preserves the finite-dimensional
		space $\ran P_L$, all moments $\cs{h_x^p}_{\sigma_{L,t}}$ are finite and differentiable in $t$, and we have, for any constant $D_p \geq 0$,
		\begin{align*}
			\cs{(h_x+D_p)^p}_{\sigma_{t}} \leq  \liminf_{L\to\infty} 	\cs{(h_x+D_p)^p}_{\sigma_{L,t}} \in [0,\infty].
		\end{align*}
		
	\end{lem}
	\begin{proof}
		The truncated dynamics preserves the finite-dimensional
		space $\ran P_L$, which is spanned by tensor products
		of Schwartz eigenfunctions. Since $\sigma_{L,t}$ is
		supported in this space, all its moments are finite
		and differentiable.
		
		The inequality
		\begin{equation}\label{eq:relative}
			\|\phi_x\phi_yu\|^2
			\leq \frac12\bigl(
			\|\phi_x^2u\|^2+\|\phi_y^2u\|^2
			\bigr) 
			\leq \frac{12}{\lambda_0}\bigl(
			\|h_x^{1/2}u\|^2+\|h_y^{1/2}u\|^2
			\bigr),
			\qquad u\in\Dom(H_{\mathrm{on}}^{1/2})
		\end{equation}
		shows that the interaction term is infinitesimally $H_{\mathrm{on}}$-bounded, and thus, $\Dom(H_\Lambda) = \Dom(H_{\mathrm{on}}) = \Dom(H^{(L)})$ for all $L$. This bound shows further that, for any $u\in\Dom(H_\Lambda)$,
		\[
		\|(H^{(L)}-H_\Lambda)u\|
		\leq\|(\Id-P_L)(H_\Lambda-H_{\mathrm{on}})u\|
		+C_\Lambda\|(H_{\mathrm{on}}+\abs{\Lambda})^{1/2}(\Id-P_L)u\|\to 0
		,
		\]
		where the last convergence follows from the fact that the projections $P_L$ commute with $H_{\mathrm{on}}$. Thus,  $H^{(L)} \to H_\Lambda$, $L \to \infty$, in the strong resolvent sense, cf. \cite[Theorem VIII.25]{RS1}. This implies the strong operator convergence of the corresponding unitary groups, and combined with the strong convergence $P_L\to\Id$ this implies $\sigma_{L,t}\to\sigma_t$ in trace norm. Thus, for all $n \in \IN$, 
		\[
		\Tr\bigl(\sigma_t\min\{(h_x+D_p)^{p},n\Id\}\bigr) =\lim_{L\to\infty}
		\Tr\bigl(\sigma_{L,t}\min\{(h_x+D_p)^{p},n\Id\}\bigr) \leq \liminf_{L\to\infty}
		\Tr\bigl(\sigma_{L,t} (h_x+D_p)^{p}\bigr) .
		\]
		Letting $n\to\infty$ shows the statement by monotone convergence.
	\end{proof}

			Finally, we can prove the main proposition of this section.
		\begin{proof}[Proof of \Cref{prop:moment}]
			Let $D$ be the constant from \Cref{lem:on-site}~\ref{it:shifted}, and consider the shifted on-site energies $\widetilde h_x =h_x+D(p/2)^2\Id$ and introduce the shifted moments
			\[
			\widetilde M_p(\sigma):=\max_{x\in\Lambda}\cs{\widetilde h_x^{p}}_\sigma .
			\]
			Write $p=2k$. We first show that
			\begin{equation}\label{eq:shifted moment bound}
				\widetilde M_p(\sigma_t)^{1/(8k)}\leq\widetilde M_p(\sigma)^{1/(8k)}+C|t|(\Tr\sigma)^{1/(8k)}
			\end{equation}
			with $C$ independent of $k$. To this end, we derive a differential inequality
			for the regularized moments $t\mapsto\Tr(\sigma_{L,t}\widetilde h_x^{2k})$ with
			constants independent of $L$ and $k$, then integrate it, and finally let
			$L\to\infty$ using \Cref{lem:spectral approx}.
			
			Write $\cs{\cdot,\cdot}_2$ for the Hilbert--Schmidt inner product. 
			Differentiating the moment, we get
			\begin{align*}
				\frac{d}{dt}\Tr(\sigma_{L,t}\widetilde h_x^{2k})
				=-\i\Tr\!\left(\sigma_{L,t}[\widetilde h_x^{2k},H^{(L)}]\right)
				=2  \Im 
				\left\langle
				\widetilde h_x^k\sigma_{L,t}^{1/2},
				[\widetilde h_x^k,H^{(L)}]\sigma_{L,t}^{1/2}
				\right\rangle_2,
			\end{align*}
			and applying
			Cauchy--Schwarz for Hilbert--Schmidt operators yields
			\begin{align}
				\label{eq:first deriv est}
				\Abs{ \frac{d}{dt}\Tr(\sigma_{L,t}\widetilde h_x^{2k}) } \leq 2 \nn{\widetilde h_x^k\sigma_{L,t}^{1/2} }_2  \nn{[\widetilde h_x^k,H^{(L)}]\sigma_{L,t}^{1/2}}_2 \leq  2\widetilde M_p(\sigma_{L,t})^{1/2} \sum_{y:\{x,y\}\in E_\Lambda}
				\|[\widetilde h_x^k,\phi_x]\phi_y\sigma_{L,t}^{1/2}\|_2,
			\end{align}
			where in the last step we used $\nn{\widetilde h_x^k\sigma_{L,t}^{1/2} }_2 \leq  \widetilde M_p(\sigma_{L,t})^{1/2}$, $[\widetilde h_x^k,H^{(L)}]
			=-\sum_{y:\{x,y\}\in E_\Lambda}
			P_L[\widetilde h_x^k,\phi_x]\phi_yP_L$, which holds since $\widetilde h_x-h_x$ is a multiple of the identity, as well as 
			$P_L\sigma_{L,t}^{1/2}=\sigma_{L,t}^{1/2}$ and
			$\|P_L\|\leq1$ to remove the cutoffs.
			
			We estimate the second factor in \eqref{eq:first deriv est} using \Cref{lem:on-site}, after writing the Hilbert--Schmidt norm in terms of an orthonormal basis of $\ran P_L$ consisting of Schwartz functions. We obtain
			\begin{align}
				\|[\widetilde h_x^k,\phi_x]\phi_y\sigma_{L,t}^{1/2}\|_2^2
				&\leq Ck^2\|\phi_y\widetilde h_x^{k-1/2}\sigma_{L,t}^{1/2}\|_2^2 \nonumber \\
				&\leq Ck^2\|\widetilde h_x^{k-1/2}\sigma_{L,t}^{1/2}\|_2
				\|\phi_y^2\widetilde h_x^{k-1/2}\sigma_{L,t}^{1/2}\|_2 \nonumber \\
				&\leq Ck^2\bigl[\Tr(\sigma_{L,t}\widetilde h_x^{2k-1})\bigr]^{1/2}
				\bigl[\Tr(\sigma_{L,t}\widetilde h_x^{2k-1}\widetilde h_y)\bigr]^{1/2}, \label{eq:hkx phix phiy est}
			\end{align}
			where we first used \Cref{lem:on-site} \ref{it:shifted}, then Cauchy--Schwarz, and finally the form inequality
			$\phi_y^4\leq(24/\lambda_0)h_y\leq(24/\lambda_0)\widetilde h_y$ together with the fact that the on-site operators at distinct sites commute.

			Both moments can be estimated using finite sums.
			Choose a common orthonormal eigenbasis
			$(\psi_j)_{j=1}^{N_L}$ of the on-site operators on $\ran P_L$, and let $e_{z,j}$ denote the corresponding eigenvalues of $\widetilde h_z$. 
			Hölder's inequality with exponents $2k$ and $2k / (2k-1)$ for finite sums then yields
			\begin{align*}
				\Tr(\sigma_{L,t}\widetilde h_x^{2k-1})
				&=\sum_j\cs{ \psi_j,\sigma_{L,t}\psi_j }e_{x,j}^{2k-1}\\
				&=\sum_j\cs{ \psi_j,\sigma_{L,t}\psi_j }^{1/(2k)}
				\bigl(\cs{ \psi_j,\sigma_{L,t}\psi_j }e_{x,j}^{2k}\bigr)^{(2k-1)/(2k)}\\
				&\leq
				\left(\sum_j\cs{ \psi_j,\sigma_{L,t}\psi_j }\right)^{1/(2k)}
				\left(\sum_j\cs{ \psi_j,\sigma_{L,t}\psi_j }e_{x,j}^{2k}\right)^{(2k-1)/(2k)} \\
				&=(\Tr\sigma_{L,t})^{1/(2k)}
				\Tr(\sigma_{L,t} \widetilde h_x^{2k} )^{1-1/(2k)}
				\\
				&\leq(\Tr\sigma_{L,t})^{1/(2k)}
				\widetilde M_p(\sigma_{L,t})^{1-1/(2k)}.
			\end{align*}
			For the mixed moment, Young's inequality for products of scalars with the same exponents yields
			\begin{align*}
				\Tr(\sigma_{L,t}\widetilde h_x^{2k-1}\widetilde h_y)
				&=\sum_j\cs{ \psi_j,\sigma_{L,t}\psi_j }e_{x,j}^{2k-1}e_{y,j}\\
				&\leq\sum_j\cs{ \psi_j,\sigma_{L,t}\psi_j }
				\left(\frac{2k-1}{2k}e_{x,j}^{2k}
				+\frac1{2k}e_{y,j}^{2k}\right)\\
				&=\frac{2k-1}{2k}\Tr(\sigma_{L,t}\widetilde h_x^{2k})
				+\frac1{2k}\Tr(\sigma_{L,t}\widetilde h_y^{2k})\\
				&\leq \widetilde M_p(\sigma_{L,t}).
			\end{align*}
			Using both estimates in \eqref{eq:hkx phix phiy est} we arrive at
			\[
			\|[\widetilde h_x^k,\phi_x]\phi_y\sigma_{L,t}^{1/2}\|_2
			\leq Ck(\Tr\sigma_{L,t})^{1/(8k)}
			\widetilde M_p(\sigma_{L,t})^{1/2-1/(8k)}.
			\]
			Substituting back into  \eqref{eq:first deriv est}, together with
			$\deg_\Lambda(x)\leq2d$, proves
			\begin{equation}
				\label{eq:reg est derivative}
				\left|\frac{d}{dt}\Tr(\sigma_{L,t}\widetilde h_x^{2k})\right|
				\leq Ck(\Tr\sigma_{L,t})^{1/(8k)}
				\widetilde M_p(\sigma_{L,t})^{1-1/(8k)}.
			\end{equation}
			
			We now use a Gronwall argument and remove the regularization.
			Since $\widetilde M_p(\sigma_{L,t})$ is the maximum of finitely
			many smooth moments, its derivative almost everywhere
			equals that of a maximizing term.
			Thus \eqref{eq:reg est derivative} and the chain rule yield
			\[
			\left|\frac{d}{dt}
			\widetilde M_p(\sigma_{L,t})^{1/(8k)}\right|
			\leq \frac{Ck}{8k}(\Tr\sigma_{L,t})^{1/(8k)}
			\leq C(\Tr\sigma)^{1/(8k)}
			\]
			for almost every $t$. The factor $k$ from \Cref{lem:on-site} \ref{it:shifted} cancels here against the factor $1/(8k)$ from the chain rule, so that $C$ does not depend on $k$.
			Here we used
			$\Tr\sigma_{L,t}=\Tr(P_L\sigma P_L)\leq\Tr\sigma$ and assumed $P_L \sigma P_L \not= 0$ (otherwise the inequality is trivial).
			Integration, together with
			$\widetilde M_p(\sigma_{L,0})\leq \widetilde M_p(\sigma)$, yields
			\[
			\widetilde M_p(\sigma_{L,t})^{1/(8k)}
			\leq \widetilde M_p(\sigma)^{1/(8k)}
			+C(\Tr\sigma)^{1/(8k)}|t|.
			\]
			Finally, \Cref{lem:spectral approx} yields $\cs{\widetilde h_x^{2k}}_{\sigma_t}\leq\liminf_{L\to\infty}\cs{\widetilde h_x^{2k}}_{\sigma_{L,t}}$ for every $x\in\Lambda$, which proves \eqref{eq:shifted moment bound}.
			
			It remains to return to $M_p$. Since $h_x$ and $\widetilde h_x$ commute and $\Id\leq h_x\leq \widetilde h_x$, we have $\css{h_x^{2k}}_{\sigma}\leq\css{\widetilde h_x^{2k}}_{\sigma}$, and $(a+b)^{2k}\leq2^{2k-1}(a^{2k}+b^{2k})$ shows that
			\[
			\css{\widetilde h_x^{2k}}_\sigma\leq 2^{2k-1}\bigl(\css{h_x^{2k}}_\sigma+(Dk^2)^{2k}\Tr\sigma\bigr).
			\]
			In particular, $\widetilde M_p(\sigma)<\infty$. Taking the $1/(8k)=1/(4p)$-th power and using \eqref{eq:shifted moment bound}, we obtain
			\[
			M_p(\sigma_t)^{1/(4p)}
			\leq\widetilde M_p(\sigma_t)^{1/(4p)}
			\leq 2^{1/4}\Bigl(M_p(\sigma)^{1/(4p)}+D^{1/4}\sqrt k\,(\Tr\sigma)^{1/(4p)}\Bigr)+C|t|(\Tr\sigma)^{1/(4p)},
			\]
			which proves the claim.
		\end{proof}

	\subsection{Approximation by a local dynamics}\label{sec:local}
	
	In this part we implement a truncation strategy, akin to \cite{KVS}. The obstruction to a standard Lieb--Robinson bound as in \cite{NRSS} is the unboundedness of the
	couplings $\phi_x\phi_y$. By \Cref{prop:moment} and Markov's inequality,
	however, the fields are large only with small probability, so cutting off
	$\phi_x$ at high field values changes the
	dynamics only slightly in the state $\sigma$. 
	To this end, for $Q \geq 1$, consider the cutoff function 
	\begin{align*}
		\chi_Q \: \IR \rightarrow [-Q,Q], \quad  \chi_Q(s) := \begin{cases}
			s &: s \in [-Q,Q], \\ Q &: s \geq Q, \\ -Q &: s \leq -Q .
		\end{cases} 
	\end{align*}
	We only cut off the couplings on
	edges which originate inside the region
 	\[\Omega =X[R] \cap \Lambda,\] 
	since using all edges of
	$\Lambda$ would produce an error proportional to $|\Lambda|$. The corresponding Hamiltonian and its time evolution will be denoted by
	\begin{align}
		\label{eq:field cutoff Hamiltonian}
		H_{\Omega,Q} :=	\sum_{x\in\Lambda}h_x	- \sum_{ \substack{\{x,y\}\in E_\Lambda \\ \{x,y\} \cap \Omega \not= \emptyset }}\chi_Q(\phi_x)\chi_Q(\phi_y) - \sum_{\{x,y\}\in E_{\Omega^c}}\phi_x\phi_y - |\Lambda| \Id,
	\end{align}
	and $\tau^{\Omega,Q}_t(A) := e^{\i t H_{\Omega,Q}} A e^{-\i t H_{\Omega,Q}}$.
	
	In the main statement of this section we quantify the error made by this cutoff. 
	\begin{lem}[Cutting off the interactions]\label{lem:clip}
		For $p\in2\IN$ and $0 \leq \sigma \in\mathfrak S_1(\cH_\Lambda)$ with
		$M_p(\sigma)<\infty$,
		\begin{equation*}
			\| (e^{-i t H_\Lambda }- e^{-i t H_{\Omega,Q} }  )  \sigma^{1/2}\|_2
			\leq \frac{C|\Omega|\,|t|}{Q^{2(p-1)}}
			\big(\sqrt{M_p(\sigma)}
			+|t|^{2p} \sqrt{\Tr \sigma}\big),
			\qquad t\in\R.
		\end{equation*}
	\end{lem}
	The factor
	$Q^{-2(p-1)}$ comes from a Markov-type estimate for large field values, see
	\eqref{eq:tail} below. The factor involving the state $\sigma$
	is the square root of the moment bound of \Cref{prop:moment}. It enters
	because the cutoff error accumulates along the trajectory $\sigma_s$,
	$0\leq s\leq t$.
	\begin{proof}
		With Duhamel we get, for $t \geq 0$,
		\begin{align}
			\label{eq:duhamel step 1}
			\|(e^{-i t H_\Lambda }- e^{-i t H_{\Omega,Q} }  ) \sigma^{1/2}\|_2
			&\leq \sum_{ \substack{\{x,y\}\in E_\Lambda \\ x \in \Omega \text{ or } y \in \Omega}}\int_0^t
			\|(\phi_x\phi_y-\chi_Q(\phi_x)\chi_Q(\phi_y)) e^{-i s H_\Lambda } \sigma^{1/2}\|_2 \d s.
		\end{align}
		Note that the application of Duhamel and differentiation with respect to the Hilbert--Schmidt norm is justified by the spectral theorem when we consider the Hamiltonians as operators acting on the Hilbert--Schmidt operator $\sigma^{1/2}$.  In fact, since $p\geq2$ and $h_x\geq1$, we have
		\[
		\|H_{\mathrm{on}}\sigma^{1/2}\|_2
		\leq \sum_{x\in\Lambda}
		\bigl(\Tr(\sigma h_x^2)\bigr)^{1/2}
		\leq |\Lambda|M_p(\sigma)^{1/2}<\infty,
		\]
		and the graph norms of $H_{\mathrm{on}}$, $H_\Lambda$ and $H_{\Omega,Q}$ are equivalent, cf. \Cref{lem:spectral approx}.

		In order to estimate the integrand in \eqref{eq:duhamel step 1},  we use joint functional calculus for the field operators $\phi_x$, $\phi_y$, which commute for $x \not= y$. 
		We now claim  $|uv-\chi_Q(u)\chi_Q(v)|^2 \leq Q^{-4(p-1)}(u^{4p}+v^{4p})$ for all $u,v\in\R$, inspired by the classical Markov inequality. The claim is trivial if $|u|,|v|\leq Q$, since the left-hand side vanishes. On the other hand, if $\abs u \geq \abs v$ and $\abs u \geq Q$,
		\begin{equation}\label{eq:tail}
			|uv-\chi_Q(u)\chi_Q(v)|^2
			\leq u^2v^2
			\leq u^4
			\leq Q^{-4(p-1)}u^{4p}
			\leq Q^{-4(p-1)}(u^{4p}+v^{4p}),
		\end{equation}
		and all other cases follow by symmetry.
		In words, the cut-off and original field products differ only where one of
		the fields exceeds $Q$, and there we trade the excess powers of the field
		for negative powers of $Q$.
		Thus, by functional calculus,
		\begin{align*}
			\|(\phi_x\phi_y-\chi_Q(\phi_x)\chi_Q(\phi_y)) e^{-i s H_\Lambda } \sigma^{1/2}\|_2^2
			\leq Q^{-4(p-1)}\Tr\big(\sigma_s(|\phi_x|^{4p}+|\phi_y|^{4p})\big).
		\end{align*}
		The moments of the field operators can be estimated with \Cref{lem:on-site} \ref{it:moment bounds}. Choosing $k=p/2$ and using $h_x\geq\Id$, this implies that $\phi_x^{2p}h_x^{-p/2}$ is bounded, uniformly in $x,\Lambda$. Thus, 
		\[
		\Tr(\sigma_s|\phi_x|^{4p})
		=\|\phi_x^{2p}\sigma_s^{1/2}\|_2^2
		\leq C\|h_x^{p/2}\sigma_s^{1/2}\|_2^2
		\leq C M_p(\sigma_s).
		\]
		
		Using the last two estimates in the Duhamel expansion \eqref{eq:duhamel step 1}, we find
		\begin{align*}
			\|(e^{-i t H_\Lambda }- e^{-i t H_{\Omega,Q} }  ) \sigma^{1/2}\|_2
			\leq C|\Omega|Q^{-2(p-1)}\int_0^t M_p(\sigma_s)^{1/2}\d s.
		\end{align*}
		Estimating the integrand with \Cref{prop:moment} proves the claim for $t \geq 0$, and negating both generators also treats $t<0$. 
	\end{proof}
	
	In the following we write
	\begin{align*}
		H^{\text{loc}}_{\Omega,Q} :=  \sum_{x\in\Omega}h_x
		-\sum_{\{x,y\}\in E_\Omega}
		\chi_Q(\phi_x)\chi_Q(\phi_y)  - \abs \Lambda \Id, \qquad \tau^{\Omega,Q,\text{loc}}_t(A) :=  e^{\i t H^{\text{loc}}_{\Omega,Q}} A e^{-\i t H^{\text{loc}}_{\Omega,Q}}, \quad A \in \cA_\Lambda,
	\end{align*}
	for the operator acting only in the region $\Omega$ with cut-off interactions. 
	
	\begin{lem}[Removing the boundary interactions]\label{lem:boundary}
		There are $C,v>0$, depending only on $d$, such that for every
		$A\in\cA_X$ with $\nn A\leq 1$ and every $t\in\R$,
		\begin{equation}\label{eq:boundary}
			\|\tau^{\Omega,Q}_t(A)-\tau^{\Omega,Q,\text{loc}}_t(A)\|
			\leq C|\Omega| Q^2|t|e^{-(R-vQ^2|t|)}.
		\end{equation}
	\end{lem}
	\begin{proof}
		The idea is to use again a Duhamel argument. To this end, consider the operator
		\begin{align}
			\label{eq:setminus partial}
			H^{\setminus \partial} := H_{\Omega,Q}
			+\sum_{\substack{\{x,y\}\in E_\Lambda\\
					x\in \Omega,\ y\notin \Omega}}
			\chi_Q(\phi_x)\chi_Q(\phi_y),
		\end{align}
		where all boundary terms are removed, so the resulting evolution on $A \in \cA_X$ agrees with that generated by $H^{\text{loc}}_{\Omega,Q}$.
		At the same time the relative unitary family $s \mapsto e^{-\i s H_{\Omega,Q}} e^{\i s H^{\setminus \partial}}$   is strongly continuously differentiable, since the two
		generators differ only by the finite sum of bounded boundary terms as written on the right-hand side of \eqref{eq:setminus partial}.  Differentiating the corresponding conjugation of $A$
		therefore yields the following Duhamel identity in the strong operator sense:
		\begin{align}
			\tau^{\Omega,Q}_t(A)- \tau^{\Omega,Q,\text{loc}}_t(A)
			&=\i\int_0^t \tau^{\Omega,Q}_{t-s}([\tau^{\Omega,Q,\text{loc}}_s(A), \sum_{\substack{\{x,y\}\in E_\Lambda\\x\in \Omega,\ y\notin \Omega}}
			\chi_Q(\phi_x)\chi_Q(\phi_y)]) \d s \nonumber \\
			&=\i \sum_{\substack{\{x,y\}\in E_\Lambda\\x\in \Omega,\ y\notin \Omega}} \int_0^t \tau^{\Omega,Q}_{t-s}([\tau^{\Omega,Q,\text{loc}}_s(A), 
			\chi_Q(\phi_x)] \chi_Q(\phi_y)) \d s, \label{eq:last duhamel line}
		\end{align}
		where we used in the second equality that $\tau^{\Omega,Q,\text{loc}}_s(A) \in \cA_\Omega$, which commutes with $\chi_Q(\phi_y)$.
		
		To bound the norm of the commutator in the integral, we use the Lieb--Robinson bound in \cite[Corollary~2.2]{NRSS} to show that there are constants $C, v_{\text{LR}}$, depending
		only on $d$, such that
		\begin{equation}\label{eq:LR}
			\|[\tau^{\Omega,Q,\text{loc}}_s(A),\chi_Q(\phi_x)]\|
			\leq CQe^{-(\dist(X,\{x\})-v_{\text{LR}} Q^2|s|)}.
		\end{equation}
		In fact, the corollary allows for unbounded self-adjoint on-site operators as in our case, 
		with decay parameter $a=1$ and
		$F_1(r)= e^{-r}(1+r)^{-d-1}$. Since the intersite interactions have range
		one, the interaction norm in the notation of \cite{NRSS}, which controls the Lieb--Robinson velocity, can be bounded by
		\begin{align*}
			\|\Phi\|_1
			&=\sup_{u,v\in\Omega}
			\frac{1}{F_1(\dist(u,v))}
			\sum_{\substack{\{z,w\}\in E_\Omega\\u,v\in\{z,w\}}}
			\|\chi_Q(\phi_z)\chi_Q(\phi_w)\| \leq C_d Q^2
		\end{align*}
		for a constant $C_d$ only depending on $d$, 
		where we used
		$\|\chi_Q(\phi_z)\|\leq Q$.
		Moreover, the support of $\chi_Q(\phi_z)$ is a singleton, so the boundary-cardinality factor in that corollary is at most one.

		This can be applied to the commutator term in \eqref{eq:last duhamel line}. For each crossing edge $\{x,y\}$ in the summation, we have $\dist(X,\{x\})=R$, and there are at most $2d|\Omega|$ such
		edges. This yields the desired statement for $t \geq 0$. 
		Again, considering the negatives of the operators proves the estimate also for negative times.
	\end{proof}
	
	\subsection{Conclusion of main result}
	We now combine the two preceding lemmas by choosing 	$A_R(t):=\tau^{\Omega,Q,\text{loc}}_t(A)$ and writing
	$\tau_t(A)-A_R(t)=\bigl(\tau_t(A)-\tau^{\Omega,Q}_t(A)\bigr)
	+\bigl(\tau^{\Omega,Q}_t(A)-\tau^{\Omega,Q,\text{loc}}_t(A)\bigr)$. The first difference is controlled
	by \Cref{lem:clip} and the second by \Cref{lem:boundary}. 
	Choosing the cutoff $Q$ is a tradeoff between the two errors. The error due to the cutoff  decreases by choosing $Q$ large enough, 
	whereas the Lieb--Robinson error is exponentially small only as long as
	$v_{\text{LR}}Q^2|t|<R$. We therefore choose $Q$ sufficiently large subject to
	$v_{\text{LR}}Q^2|t|\leq R/2$, namely $Q^2=R/(2v_{\text{LR}}(1+|t|))$.
	\begin{proof}[Proof of \Cref{thm:main}]
		\begin{enumerate}[label=(\alph*),wide,nosep]
			\item 		
			With standard Schatten norm inequalities, Cauchy--Schwarz for Schatten norms,  and finally \Cref{lem:clip} we obtain, for all $Q \geq 1$,
			\begin{align}
				\big|\Tr\big(\sigma(\tau_t(A)-\tau^{\Omega,Q}_t(A))\big)\big|
				&\leq \nn{A}
				\|\tau_{-t}(\sigma)-\tau^{\Omega,Q}_{-t}(\sigma)\|_1
				\notag\\
				&\leq
				2(\Tr\sigma)^{1/2}
				\|(e^{-\i t H_\Lambda}-e^{-\i t H_{\Omega,Q}})\sigma^{1/2}\|_2
				\notag\\
				&\leq
				C|\Omega|\sqrt{M_p(\sigma)}
				\frac{|t|(1+|t|)^{2p}}{Q^{2(p-1)}}.
				\label{eq:expectclip}
			\end{align}
			\Cref{lem:boundary} yields
			\begin{align}
				\label{eq:lem_boundary application}
				\|\tau^{\Omega,Q}_t(A)-\tau^{\Omega,Q,\text{loc}}_t(A)\|
				\leq C|\Omega|Q^2|t|e^{-(R-v_{\text{LR}}Q^2|t|)}.
			\end{align}
			We now choose $Q^2 := \frac{R}{2v_{\text{LR}}(1+|t|)}$ and $C_0 := 2v_{\text{LR}}$, so that $Q \geq 1$ for $R \geq C_0(1+\abs t)$. Combine \eqref{eq:lem_boundary application} with \eqref{eq:expectclip}, then  use $|\Omega|\leq |X|(2R+1)^d\leq C_d |X|R^d$   to get
			\begin{align*}
				\big|\Tr\big(\sigma(\tau_t(A)-\tau^{\Omega,Q,\text{loc}}_t(A))\big)\big| &\leq C|\Omega|\sqrt{M_p(\sigma)}\,|t|
				\left(
				\frac{(1+|t|)^{2p}}{Q^{2(p-1)}}
				+Q^2 e^{-(R-v_{\text{LR}}Q^2|t|)}
				\right) \\
				&\leq C \abs X R^d  \sqrt{M_p(\sigma)}\,|t| 	\left(
				\frac{(1+|t|)^{3p-1}}{R^{p-1}} +
				\frac{R e^{-R/2}}{1+|t|}
				\right)  \\
				&= C|X|\sqrt{M_p(\sigma)}\,
				\frac{|t|(1+|t|)^{3p-1}}{R^{p-d-1}}	\left(
				1 +
				\frac{R^p e^{-R/2}}{(1+|t|)^{3p}}
				\right). 
			\end{align*}
			This together with $\sup_{R\geq1}R^pe^{-R/2}<\infty$ shows \ref{it:localexpect}, when we set $A_R(t) := \tau^{\Omega,Q,\text{loc}}_t(A)$.
			Note that this choice is the same for all $\Lambda \supseteq X[R+1]$, as $H^{\text{loc}}_{\Omega,Q}$ involves only the on-site energies $h_x$, $x\in\Omega$, and couplings on edges inside $\Omega$. The cutoff $Q$  depends  only on $R$ and $t$. The operators $h_x$ contain the vertex degree inside $\Lambda$, but for $\Lambda \supseteq X[R+1]$, every $x\in\Omega$ has all its $2d$ neighbors in $\Lambda$. The remaining term $-|\Lambda|\Id$ does not affect the dynamics induced by this Hamiltonian.
			
			\item
			In part~\ref{it:localexpect}, the dynamics could be moved entirely
			onto the state. Now $\sigma^{1/2}$ acts only from the right, so the
			difference of the two dynamics has to be split into a part acting before
			$A$ and a part acting after $A$. In the second part, the state
			$A\sigma_tA^*$ appears, and controlling its moments is the only place
			where \eqref{eq:preserve} is used.
			We start with the identity
			\[
			\left( \tau_t(A)-\tau^{\Omega,Q}_t(A) \right) \sigma^{1/2}
			=
			e^{\i t H_{\Omega,Q}}A(e^{-\i t H_\Lambda}-e^{-\i t H_{\Omega,Q}}) \sigma^{1/2}
			+(e^{\i t H_\Lambda}-e^{\i t H_{\Omega,Q}})Ae^{-\i t H_\Lambda} \sigma^{1/2}.
			\]
			The first term can be bounded in Hilbert--Schmidt norm with \Cref{lem:clip}
			by the same expression \eqref{eq:expectclip}.
			For the second one, we apply  \Cref{lem:clip} with
			$\sigma$ replaced by $A \sigma_t A^*$ and time $-t$,
			\begin{align}
				\|(e^{\i t H_\Lambda}-e^{\i t H_{\Omega,Q}})Ae^{-\i t H_\Lambda}\sigma^{1/2}\|_2
				&=
				\|(e^{\i t H_\Lambda}-e^{\i t H_{\Omega,Q}})(A\sigma_tA^*)^{1/2}\|_2 \nonumber \\
				&\leq 
				\frac{C|\Omega|\,|t|}{Q^{2(p-1)}}
				\left(\sqrt{M_p(A\sigma_tA^*)}
				+|t|^{2p} \sqrt{ \Tr(A \sigma_t A^*) }\right). \label{eq:Utstar - Vtstar}
			\end{align}
			We have $\Tr(A \sigma_t A^*) \leq \Tr \sigma \leq M_p(\sigma)$, and with \eqref{eq:preserve} and
			\Cref{prop:moment}  we can bound
			\begin{align*}
				\sqrt{M_p(A\sigma_tA^*)} 
				\leq\kappa \sqrt{M_p(\sigma_t )} 
				\leq C \kappa \sqrt{M_p(\sigma)}
				\bigl(1
				+|t|^{2p}\bigr).
			\end{align*}
			Inserting these bounds into \eqref{eq:Utstar - Vtstar} yields the same upper bound as on the right-hand side of \eqref{eq:expectclip} (up to a factor $\kappa$). Combining this again with \eqref{eq:lem_boundary application} proves the statement. \qedhere
		\end{enumerate}
	\end{proof}

	\section{Proof of digital quantum simulation result}\label{sec:simulation}
	
	For the proof of the simulation result we proceed in two steps. First in \Cref{sim:lem:truncation} we provide  an error estimate for the approximation of the full dynamics by the truncated dynamics where we restrict to a finite number of harmonic levels. In the second step in \Cref{sim:lem:circuit}, we show how the truncated dynamics can be in turn approximated by a quantum circuit. 
	
	We state the two estimates and first show how they imply \Cref{sim:thm:main}, while their proofs follow in separate subsections. Throughout, constants $C$ can change from line to line and only depend on $d,m_0,\lambda_0$.
	
	For $K\in\N_0$, define the truncated Hamiltonian and its propagator by
	\begin{equation}\label{sim:eq:HK}
		H_K=\Pi_KH_\Lambda\Pi_K,\qquad
		U_K(t)=e^{-\i tH_K}.
	\end{equation}
	For a site $x$, we can introduce the creation and annihilation operators $a_x^* = (\phi_x-\i\pi_x) / \sqrt2$ and $a_x = (\phi_x+\i\pi_x) / \sqrt2$, which raise and lower the occupation numbers by one, respectively. 
	In particular, they satisfy the standard norm estimates
	\begin{equation}\label{eq:ladder norms}
		\|a_x\Pi_{x,K}\|=\sqrt K,
		\qquad
		\|a_x^*\Pi_{x,K}\|=\sqrt{K+1}.
	\end{equation}
	Analogously to the definition of $M_p(\sigma)$ we also introduce the notation
	\begin{align*}
		M_p(u) := M_p(\ketbra{u}{u}) = \max_{x\in\Lambda} \nn{ h_x^{p/2} u }^2. 
	\end{align*}
	The first lemma bounds the error when we restrict to the first $K+1$ harmonic levels. We will write $u_s := e^{-\i s H_\Lambda} u$, $s \in \IR$, for time-evolved vector states.
\begin{lem}[Harmonic truncation]\label{sim:lem:truncation}
	There exist constants $C\geq1$ and $c_0>0$, depending only on
	$d,m_0,\lambda_0$, such that for all $T\geq1$ and
	$n_0,K\in\N_0$ satisfying $K+1\geq C(n_0^2+T^4)$, one has
	\begin{equation}\label{sim:eq:truncation}
		\bigl\|(e^{-\i T H_\Lambda}-U_K(T))\Pi_{n_0}\bigr\|
		\leq CNT\,e^{-c_0\sqrt{K+1}}.
	\end{equation}
\end{lem}

	The second lemma concerns the approximation of the finite-dimensional propagator by a sequence of unitary 1- and 2-qubit gates. The idea is standard (cf. \cite{HHKL,BCCKS2017})
	but we need to track the dependence on $K$, similarly to \cite{KVS}. 	Assume the computability condition on $m_0^2$ and $\lambda_0$ as in \Cref{sec:simulation results}.
	\begin{lem}[Finite simulation circuit]\label{sim:lem:circuit}
		Let $K+1=2^q$ with $q\in\IN$,
		$T\ge1$, and $\eta\in(0,1)$. Then there exist an integer number of ancilla qubits $a \geq 0$
		and a unitary circuit $U$, consisting of $G$ one- and two-qubit
		gates, acting on
		\[
		\cH_K\otimes(\mathbb C^2)^{\otimes a}
		\cong
		(\mathbb C^2)^{\otimes Nq}
		\otimes(\mathbb C^2)^{\otimes a},
		\]
		such that, for every density matrix $\rho$ on $\cH_K$,
		\[
		\left\|
		\Tr_{\mathrm{anc}}\!\left(
		U\bigl(\rho\otimes
		|0\rangle\langle0|^{\otimes a}\bigr)U^*
		\right)
		-U_K(T)\rho U_K(T)^*
		\right\|_1
		\le\eta.
		\]
		The number of ancilla qubits $a$ and gates $G$ are bounded by
		\begin{equation}\label{sim:eq:finitecost}
			a\le C\log^c\!\frac{2NT(K+1)}{\eta}, \qquad G \leq
			CNT(K+1)^2\log^c\!\frac{2NT(K+1)}{\eta},
		\end{equation}
		where $c \geq 1$ is an integer depending only on $d$.
	\end{lem}

	We now combine the lemmas. Increasing~$K$ reduces the truncation error but increases the circuit cost.
	\begin{proof}[Proof of \Cref{sim:thm:main}]
		Choose $K+1=2^q$ to be the smallest power of two satisfying
		\begin{equation}\label{sim:eq:cutoff}
			K+1\ge C\left(n_0^2+T^4+\log^2\frac{2NT}{\varepsilon}\right)\,.
		\end{equation}
		For $C$ sufficiently large this ensures $K\ge n_0$ and makes the error in \eqref{sim:eq:truncation} at most $\varepsilon/4$. Thus, with $\rho=\Pi_{n_0}\rho\Pi_{n_0}$, we obtain
		\begin{equation*}
			\|e^{-\i T H_\Lambda}\rho e^{\i T H_\Lambda}
			-U_K(T)\rho U_K(T)^*\|_1 \leq 2 \nn{ (e^{-\i T H_\Lambda}-U_K(T))\Pi_{n_0} } \leq \varepsilon / 2. 
		\end{equation*}
		In order to approximate $U_K(T)\rho U_K(T)^*$ with a quantum circuit up to an error $\varepsilon / 2$,  we apply \Cref{sim:lem:circuit} with that choice of $K$ and $\eta=\varepsilon/2$. This shows \eqref{sim:eq:error}.

Rounding up to the next power of two in \eqref{sim:eq:cutoff} yields
$K+1\leq C(n_0^2+T^4+L^2)\leq C(n_0TL)^4$, since $n_0,T,L\geq1$.
Thus, $q=\log(K+1)\leq C\log(2n_0TL)$. Squaring the first bound gives
$(K+1)^2\leq C(n_0^4+T^8+L^4)$. Since
$\log(4NT(K+1)/\varepsilon)=\log(4NT/\varepsilon)+q\leq CL$,
\eqref{sim:eq:finitecost} with $\eta=\varepsilon/2$ shows the stated
ancilla and gate bounds.
	\end{proof}
	
	\subsection{Error from harmonic truncation}
	
	In order to prove the error bound, we first have to translate the propagation bounds for the energy moments to propagation bounds for the harmonic occupation numbers.
		\begin{lem}[Harmonic occupation moments]\label{sim:lem:uniformoccupation}
		For every $k\in\IN$, $n_0\in\N_0$, unit vector $u\in\ran\Pi_{n_0}$ and $t\in\R$,
		\begin{equation*}
			\max_{x\in\Lambda}\|(n_x+\Id)^k u_t\|
			\leq \bigl[C(n_0^2+k^2+t^4)\bigr]^k\,,
		\end{equation*}
		where $C$ depends only on $d,m_0,\lambda_0$.
	\end{lem}
	\begin{proof}
		Let $p:=2k$. Since $h_x-\Id$ is a polynomial of degree four in $a_x$ and $a_x^*$, it raises the occupation number by at most four, so 	$h_x\Pi_{x,m}=\Pi_{x,m+4}\,h_x\Pi_{x,m}$,
	and \eqref{eq:ladder norms} implies $
		\|h_x\Pi_{x,m}\|\leq C(m+1)^2$ for all
		$m\in\N_0$.
		As $u\in\ran\Pi_{x,n_0}$ for every $x$, iterating this yields
		\[
		M_p(u)^{1/2}
		\leq\prod_{j=0}^{k-1}C(n_0+4j+1)^2
		\leq\bigl[C(n_0+k+1)^2\bigr]^k.
		\]
		Then we can apply \Cref{prop:moment} to $\ketbra{u}{u}$ to obtain
		\[
		M_p(u_t)^{1/(4p)}
		\leq C\Bigl(\sqrt{n_0+k+1}+\sqrt{k}+|t|\Bigr)
		\leq C\bigl(\sqrt{n_0+k+1}+|t|\bigr),
		\]
		and therefore $\|h_x^ku_t\|\leq M_p(u_t)^{1/2}\leq\bigl[C(\sqrt{n_0+k+1}+|t|)\bigr]^{4k}\leq\bigl[C(n_0+1+k+t^2)^2\bigr]^k$ for every $x$. Finally, we combine this with \Cref{lem:on-site} \ref{it:moment bounds} choosing $w = n_x+\Id=\tfrac12(\pi_x^2+\phi_x^2+\Id)$ and get
		\[
		\max_{x\in\Lambda}\|(n_x+\Id)^ku_t\|
		\leq C^k\Bigl(\bigl[C(n_0+1+k+t^2)^2\bigr]^k+k^{2k}\Bigr).
		\]
		Since $k\geq1$, after enlarging $C$, this implies the assertion.
	\end{proof}
	
	\begin{proof}[Proof of \Cref{sim:lem:truncation}]
		Let $u\in\ran\Pi_{n_0}$ be normalized. 
		We start by estimating
		\begin{align*}
			\|u_T-U_K(T)u\|\le\|(\Id-\Pi_K)u_T\|+ \|\Pi_K u_T - U_K(T)u\|.
		\end{align*}
		For the first term, we can use $\Id-\Pi_K\leq\sum_x(\Id-\Pi_{x,K})$ to decompose it into local tails. For the second one, we first apply Duhamel  by differentiating $U_K(T-s)\Pi_Ku_s$ and integrating,
		\begin{equation*}
			\Pi_Ku_T - U_K(T)u = - \i\int_0^T U_K(T-s)\Pi_KH_\Lambda(\Id-\Pi_K)u_s \d s,
		\end{equation*}
		where we used that $\ran \Pi_K \subset \Dom(H_\Lambda)$. 
		Thus, we arrive at
		\begin{align}
			\label{eq:two terms to control}
			\|u_T-U_K(T)u\|\le	\left( \sum_{x\in\Lambda} \nn{(\Id-\Pi_{x,K}) u_T}^2 \right)^{1/2}+\int_0^T\|\Pi_KH_\Lambda(\Id-\Pi_K)u_s\| \d s.
		\end{align}
		
		Next, in order to express the integral term in \eqref{eq:two terms to control} with local tails as well, we first expand the Hamiltonian $H_\Lambda$ as a sum over its local terms and note that 
		\[
		\Pi_K(h_x-\Id)(\Id-\Pi_K)=\Pi_K(h_x-\Id)(\Id-\Pi_{x,K}),\qquad
		\Pi_K\phi_x\phi_y(\Id-\Pi_K)=\Pi_K\phi_x\phi_y(\Id-\Pi_{x,K}\Pi_{y,K}).
		\]
		Each of those terms can be estimated with the ladder operator norm estimates \eqref{eq:ladder norms} by
		\begin{equation}\label{sim:eq:localnorm}
			\|(h_x-\Id)\Pi_{x,K}\|\le C(K+1)^2,\qquad
			\|\phi_x\phi_y\Pi_{x,K}\Pi_{y,K}\|\le C(K+1)^2.
		\end{equation}
		Thus using the triangle inequality over the sum of all local terms, 
		\begin{align}
			\nonumber
			\|\Pi_KH_\Lambda &(\Id-\Pi_K)u_s\|
			\\ & \le C(K+1)^2  \left[
			\sum_{x\in\Lambda}\|(\Id-\Pi_{x,K})u_s\|
			+\sum_{\{x,y\}\in E_\Lambda}
			\bigl(\|(\Id-\Pi_{x,K})u_s\|+\|(\Id-\Pi_{y,K})u_s\|\bigr)
			\right]. 			\label{sim:eq:boundary}
		\end{align}
				By the spectral theorem and \Cref{sim:lem:uniformoccupation}, we find for every $k \in \IN$,
		\begin{align}
			\label{eq:sup max tails}
			\sup_{0\le s\leq T}\max_{x\in\Lambda}\|(\Id-\Pi_{x,K})u_s\|
			\le \left[\frac{C_*(n_0^2+k^2+T^4)}{K+1}\right]^k,
		\end{align}
		 where $C_*$ denotes the constant in
		\Cref{sim:lem:uniformoccupation}. Choose
$k=\lfloor\sqrt{(K+1)/(4C_*)}\rfloor$. Assuming that the constant $C$ in the
assumption $K+1\geq C(n_0^2+T^4)$ satisfies $C \geq 4C_*$ ensures that
$k\geq1$ and that both $C_*(n_0^2+T^4)$ and $C_*k^2$ are at most
$(K+1)/4$. Hence, \eqref{eq:sup max tails} can be bounded by $2^{-k} \leq 2e^{-c_0\sqrt{K+1}}$ for some $c_0 >0$. 

By \eqref{eq:sup max tails}, the first term in \eqref{eq:two terms to control} is at most $2\sqrt{N} e^{-c_0\sqrt{K+1}}$. Since each site occurs at most
$1+2d$ times in \eqref{sim:eq:boundary}, the integral term  in \eqref{eq:two terms to control} is at most
$CNT(K+1)^2e^{-c_0\sqrt{K+1}}$. The prefactor $(K+1)^2$ can be absorbed into the exponential at the cost of decreasing $c_0>0$. 
	\end{proof}	
	\subsection{Finite-dimensional simulation}
	We use the strategy of Kuwahara--Vu--Saito \cite[Eqs. (34)--(45)]{KVS} because it keeps explicit control of the simulation cost as the occupation number and the local interaction norms grow. It combines the spatial decomposition of \cite{HHKL} with simulation of sparse matrices. We apply it below to the truncated $\phi^4$ Hamiltonian.
	
	\begin{proof}[Proof of Lemma~\ref{sim:lem:circuit}]
		The proof consists of three steps. We first approximate the local matrix
		entries by finite binary representations, then decompose the
		resulting evolution into smaller spatial blocks and simulate each block
		by a quantum circuit. 
		
		\medskip
		\noindent
		\textbf{Step 1: Binary representation and approximation of the entries.}
		We construct a self-adjoint approximation~$H_K^{(b)}$ with
		entries in binary encoding such that its evolution differs from $U_K(T)$ by at most $\eta$.
		
		On the encoded space
		$\bigotimes_{x\in\Lambda}\mathbb C^{2^q}$, write
		$H_K=\sum_ZH_{K,Z}$, for the finite local matrices
		\begin{equation}
			\label{eq:finite matrices}
			H_{K,\{x\}}=\Pi_{x,K}(h_x-\Id)\Pi_{x,K},\qquad
			H_{K,\{x,y\}}
			=-(\Pi_{x,K}\phi_x\Pi_{x,K})
			\otimes(\Pi_{y,K}\phi_y\Pi_{y,K}).
		\end{equation}
		For $b\in\IN$, we approximate each allowed matrix entry
		by a multiple of $2^{-b}$ with absolute error at most
		$2^{-b}$, assigning conjugate values to transposed
		entries and keeping entries outside the allowed
		occupation shifts equal to zero.
		Then we replace each matrix entry by its binary encoding.
		Denote the resulting self-adjoint matrices by $H_{K,Z}^{(b)}$
		and set $H_K^{(b)}=\sum_ZH_{K,Z}^{(b)}$.
		
		Each entry of the matrices in \eqref{eq:finite matrices} is a polynomial in $m_0^2$, $\lambda_0$, $\deg_\Lambda(x)$ and the occupation numbers, multiplied by the square root of an integer at most $(K+4)^4$. The $(b+\cO(q))$-bit approximations of $m_0^2$ and $\lambda_0$ are computed once classically, using the assumption of \Cref{sim:thm:main}, and enter the circuit as fixed constants. All remaining operations are additions, multiplications and integer square roots (via Newton iteration) of $\cO(q+b)$-bit numbers. Hence the indices and values to accuracy $2^{-b}$ can be computed with a number of gates polynomial in $q+b$, with a degree independent of $m_0$ and $\lambda_0$.
		
		Expressing $\phi$ and $\pi$ in terms of creation and annihilation operators, one sees that the on-site terms in \eqref{eq:finite matrices} can only change the occupation number by $0,\pm2,\pm4$ (five possibilities), whereas the inter-site terms can only change each vertex of one edge by $\pm 1$, respectively (four possibilities). Thus, the rows of the matrices in \eqref{eq:finite matrices} have at most five and four non-zero entries, respectively, and we get
		\begin{equation*}
			\|H_K-H_K^{(b)}\|
			\le (5N+4|E_\Lambda|)\,2^{-b}
			\le CN2^{-b}.
		\end{equation*}
		By Duhamel's formula we obtain
		\begin{equation*}
			\|e^{-iTH_K}-e^{-iTH_K^{(b)}}\|
			\le T\|H_K-H_K^{(b)}\|
			\le CNT2^{-b}\le\eta
		\end{equation*}
		when choosing $b=\lceil\log_2(CNT/\eta)\rceil$, with~$C$
		sufficiently large.

		\noindent
		\textbf{Step 2: Spatial decomposition.}
		We approximate $e^{-iTH_K^{(b)}}$ by a product of evolutions
		under Hamiltonians restricted to small spatial blocks.
		For every $\ell\in\IN$ and $m \in \IN$, divide the time interval $[0,T]$ into $m$ intervals of length $\Delta t = T / m$. The construction in~\cite{HHKL}, as used in
		\cite[Eqs.~(40)--(43)]{KVS}, produces local block Hamiltonians with corresponding time evolution
		\begin{equation*}
			H_{K,B_j}^{(b)}
			:=\sum_{Z\subseteq B_j}H_{K,Z}^{(b)}, 	\qquad	V_j:=e^{ \i \sigma_j \Delta t H_{K,B_j}^{(b)}},
		\end{equation*}
		where each $B_j\subseteq\Lambda$ is of diameter~$O(\ell)$ for $j=1, \ldots, r$ with  $r\le CNm$ and  $\sigma_j \in \{\pm 1\}$. Note that each $V_j$ performs either a forward or backward time evolution. According to \cite[Eq. (43)]{KVS} their ordered product
		approximates the actual time evolution as
		\begin{align}
			\label{eq:kvs decomposition}
			\|e^{-iTH_K^{(b)}}-V_r\cdots V_1\|
			\le CN m
			e^{-\mu\ell+vg\Delta t},
		\end{align}
		where $C,\mu,v>0$ are constants depending only on the lattice dimension. The maximal local interaction strength $g$ satisfies
		\[
		g:=\max_{x\in\Lambda}\sum_{Z\ni x}\|H_{K,Z}^{(b)}\|
		\le C_1(K+1)^2,
		\]
		by \eqref{sim:eq:localnorm}, the bounded lattice degree,
		and the rounding estimate from Step 1.
		
		Choose now  $m := \lceil C_1(K+1)^2 T\rceil$ and 
		$
		\ell=\left\lceil
		C_2\log\frac{NT(K+1)}{\eta}
		\right\rceil.
		$
		Using these choices in \eqref{eq:kvs decomposition} and choosing $C_2 >0$ large enough shows
		\begin{align*}
			\|e^{-iTH_K^{(b)}}-V_r\cdots V_1\|\le\eta.
		\end{align*}

		\medskip
		\noindent
		\textbf{Step 3: Simulation of the blocks.}
		Using the methods of \cite{KVS} we show that there is a simulation $\widetilde V_j$ of each block evolution $V_j$ 
		such that, for some number of ancilla qubits $a \geq 0$, 
		\begin{equation}\label{eq:Vj simulation}
			\|\widetilde V_jJ-JV_j\|\le \frac{\eta}{r},
			\qquad
			Ju:=u\otimes|0\rangle^{\otimes a},
		\end{equation}
		where all operators are extended by the identity outside
		the block. We use a common ancilla register large enough
		for every block circuit.
		
		Each block acts on $q|B_j|$ qubits, has row sparsity
		$s_j\le C|B_j|$, and satisfies
		$\|H_{K,B_j}^{(b)}\|_{\max}\le g|B_j|$.
		In \cite[Eq.~(38)]{KVS}, we therefore take
		$L=B_j$, $d_L=s_j$, $\tau=\Delta t$, and replace
		$|L|\log_2(\bar q+1)$ by $q|B_j|$.
		Since the Hamiltonian is time independent and
		$g\Delta t\le1$, the parameters satisfy
		\[
		\widetilde\tau=s_j^2g|B_j|\Delta t\le C|B_j|^3,
		\qquad
		\widetilde\tau'=0.
		\]
		
		For the complexity estimates, we use the upper bound
		$\widehat\tau=C_*|B_j|^3$, with $C_*>0$ sufficiently large (such that $\log \log C_* > 0$).
		For a given error $\epsilon_{\text{KVS}} \in (0,1)$,  \cite[Eq. (38)]{KVS} and the underlying sparse simulation theorem \cite[Theorem~2.1]{BCCKS2017}  bound the additional gate count and the number of matrix
		queries by
		\[
		Cq|B_j|\,\widehat\tau
		\frac{\log^2(\widehat\tau/\epsilon_{\text{KVS}})}
		{\log\log(\widehat\tau/\epsilon_{\text{KVS}})}, \quad \text{and} \qquad 
		C\widehat\tau
		\frac{\log(\widehat\tau/\epsilon_{\text{KVS}})}
		{\log\log(\widehat\tau/\epsilon_{\text{KVS}})},
		\]
		respectively. To simplify these expressions, note that $\log\log(\widehat\tau/\epsilon_{\text{KVS}})\ge\log\log C_*>0$ and
		\[
		\log(\widehat\tau/\epsilon_{\text{KVS}})
		=\log C_*+3\log|B_j|+\log(1/\epsilon_{\text{KVS}})
		\le C\log\!\left(\frac{2|B_j|}{\epsilon_{\text{KVS}}}\right).
		\]
		Thus, after enlarging $C$, the sum of the additional gates
		and the matrix queries, counted at unit cost, is at most
		\[
		\cO\left(|B_j|^4q\log^2\!\left(\frac{2|B_j|}{\epsilon_{\text{KVS}}}\right)\right).
		\]
		
		We next include the cost of implementing these queries.
		The positions and values of the nonzero entries are
		computed by enumerating the $\cO(|B_j|)$ local terms and
		combining their contributions. By Step~1, these
		computations can be implemented with a number
		of gates polynomial in $|B_j|,q,b$.
		Thus, the full
		gate count to simulate one $V_j$ is of order
		\[
		\cO \left(
		\text{poly}(|B_j|,q,b) \log^2\!\left(\frac{2|B_j|}{\epsilon_{\text{KVS}}}\right) \right).
		\]
		
		In order to get an error $\eta$ for all $r$ blocks, choose $\epsilon_{\text{KVS}}=\eta/r$.
		Since $|B_j|\le C\ell^d$, $r \leq C N(K+1)^2T$, while
		$q,b,\ell,\log(r/\eta)$ are bounded by
		$C\log(NT(K+1)/\eta)$, the preceding estimate yields a total gate count as given in \eqref{sim:eq:finitecost}. 
		
		The lattice registers use $Nq$ qubits. If the $j$-th circuit contains $G_j$ gates, it uses at most $2G_j$ ancillary qubits, since each gate acts on at most two qubits. All block circuits can use a common register
		of size $a:=\max_j a_j$. This shows the claimed qubit bounds.
		
		\medskip
		\noindent
		\textbf{Conclusion.}
		Set $U:=\widetilde V_r\cdots\widetilde V_1$. Then telescoping combined with \eqref{eq:Vj simulation} shows that
		\begin{align*}
			\|UJ-JV_r\cdots V_1\|
			&=
			\left\|
			\sum_{j=1}^r
			\widetilde V_r\cdots\widetilde V_{j+1}
			(\widetilde V_jJ-JV_j)
			V_{j-1}\cdots V_1
			\right\| \le r\epsilon_{\text{KVS}}=\eta.
		\end{align*}
		Together with Steps~1 and~2, this implies
		$
		\|UJ-JU_K(T)\|\le3\eta$. 
		Hence, for every density matrix $\rho$,
		\[
		\left\|
		\Tr_{\mathrm{anc}}(UJ\rho J^*U^*)
		-U_K(T)\rho U_K(T)^*
		\right\|_1 = 	\left\|
		\Tr_{\mathrm{anc}}(UJ\rho J^*U^*
		-J U_K(T)\rho U_K(T)^* J^* )
		\right\|_1
		\le6\eta.
		\]
		Choosing $\eta/6$ in place of
		$\eta$ proves the asserted error bound without changing
		the resource estimates.
	\end{proof}

	\section{Proof of thermodynamic limit result}\label{app:thermodynamic}
	
	The existence of the thermodynamic limit as well as its continuity are standard and an immediate consequence of the local approximation result \Cref{thm:main}, in which we can choose the local approximant independently of the finite volume $\Lambda$. 
	\begin{proof}[Proof of \Cref{cor:thermodynamic}]
		First, consider a local observable $A\in\cA_X$ with $\nn A\leq1$ for a finite set $\emptyset \not= X\subset\mathbb Z^d$.  For $R\in\IN$ with $R\geq C_0(1+|t|)$ and all finite $\Lambda\supseteq X[R+1]$, we  apply \Cref{thm:main}~\ref{it:localexpect} with  $\sigma:=\rho_\Lambda$ being the corresponding density matrix of $\omega$ on $\cH_\Lambda$, and obtain
		\[
		\big|\omega\big(\tau_t^\Lambda(A)\big)-\omega\big(A_R(t)\big)\big|
		\leq C|X|\sqrt{M_p(\omega)}\,\frac{|t|(1+|t|)^{3p-1}}{R^{p-d-1}}
		\qquad\text{for all finite }\Lambda\supseteq X[R+1],
		\]
		for $A_R(t)\in\cA_{X[R]}$ independently of $\Lambda$.
		Since $p>d+1$, the right-hand side tends to zero as $R\to\infty$. Hence the net $\big(\omega(\tau_t^\Lambda(A))\big)_\Lambda$ is Cauchy and converges. 	For general quasi-local $A\in\cA$, the result follows with the approximation by local observables and the fact that $\tau_t^\Lambda$ is isometric. The limit is a state because positivity and normalization are conserved in the pointwise limit.
		
		The continuity for local observables $A$ can be obtained from the continuity of the map $t\mapsto\omega(\tau_t^\Lambda(A))$ for fixed volumes $\Lambda$ and the convergence $\omega(\tau_t^\Lambda(A))\to\omega_t(A)$ uniformly for $t$ in compact sets. For general $A\in\cA$, the claim follows again by approximation and $|\omega_t(A)-\omega_t(B)|\leq\|A-B\|$.
	\end{proof}

	\subsection*{AI Declaration}
	The main proof idea for the propagation bound came from the authors. The authors  acknowledge the use of Artificial Intelligence to propose specific estimates and to assist in preparing an initial draft, which they then substantially rewrote to improve the structure and presentation. The AI help was especially valuable for the application to quantum simulation, a topic which did not originally lie within either author's area of expertise. The authors take  responsibility for the entire content.

	\subsection*{Acknowledgments}
	The research of ML is supported by the Deutsche Forschungsgemeinschaft (DFG, German Research Foundation) through the grants TRR 352 – Project-ID 470903074 and FOR 5413 – Project-ID 465199066, as well as by the European Union through ERC Starting Grant MathQuantProp, Grant Agreement 101163620.\footnote{Views and opinions expressed are however those of the authors only and do not necessarily
		reflect those of the European Union or the European Research Council Executive Agency. Neither
		the European Union nor the granting authority can be held responsible for them.} OS thanks Martin Fraas for a discussion that led to an improvement in the quantum simulation result.
	
	\printbibliography
\end{document}